\documentclass[sigconf]{acmart}
\usepackage{multirow}
\usepackage{colortbl}
\usepackage{soul}
\usepackage{float}
\AtBeginDocument{%
  }

\copyrightyear{2025}
\acmYear{2025}
\setcopyright{cc}
\setcctype{by}
\acmConference[CHI '25]{CHI Conference on Human Factors in Computing Systems}{April 26-May 1, 2025}{Yokohama, Japan}
\acmBooktitle{CHI Conference on Human Factors in Computing Systems (CHI '25), April 26-May 1, 2025, Yokohama, Japan}\acmDOI{10.1145/3706598.3714287}
\acmISBN{979-8-4007-1394-1/25/04}

\begin{document}

\title[Exploring Autistic Individuals' Visions for AI in Coping with Negative Self-Talk]{Reimagining Support: Exploring Autistic Individuals' Visions for AI in Coping with Negative Self-Talk}

\author{Buse Carik}

\orcid{1234-5678-9012}
\affiliation{%
  \institution{Department of Computer Science, Virginia Tech}
  \city{Blacksburg}
  \state{Virginia}
  \country{USA}
}
\email{buse@vt.edu}

\author{Victoria Izaac}
\affiliation{%
  \institution{Department of Psychology, \\Virginia Tech}
  \city{Blacksburg}
  \state{Virginia}
  \country{USA}
}
\email{victoriaizaac@vt.edu}

\author{Xiaohan Ding}
\affiliation{%
  \institution{Department of Computer Science, Virginia Tech}
  \city{Blacksburg}
  \state{Virginia}
  \country{USA}
}
\email{xiaohan@vt.edu}

\author{Angela Scarpa}
\affiliation{%
  \institution{Department of Psychology, \\Virginia Tech}
  \city{Blacksburg}
  \state{Virginia}
  \country{USA}
}
\email{ascarpa@vt.edu}

\author{Eugenia Rho}
\affiliation{%
  \institution{Department of Computer Science, Virginia Tech}
  \city{Blacksburg}
  \state{Virginia}
  \country{USA}
}
\email{eugenia@vt.edu}

\renewcommand{\shortauthors}{Carik et al.}
\definecolor{b_skyblue}{rgb}{0.1098, 0.5686, 0.7411}
\definecolor{b_orange}{rgb}{0.9450, 0.6117, 0}
\definecolor{b_plum}{rgb}{0.7, 0.3, 0.5}
\definecolor{b_green}{rgb}{0.0274, 0.4274, 0.2352}
\definecolor{b_mustard}{rgb}{0.8, 0.7, 0.2} 
\definecolor{b_red}{rgb}{0.8392, 0.3921, 0.2666}
\definecolor{b_purple}{rgb}{0.6, 0.4, 0.7} 
\definecolor{b_seafoam}{rgb}{0.5, 0.75, 0.6}
\definecolor{b_blue}{rgb}{0.3333, 0.3764, 0.6627}

\newcommand{\textft}[1]{#1}
\newcommand{\customul}[2][black]
{\setulcolor{#1}\ul{#2}\setulcolor{black}}

% Text Color
\newcommand{\TSt}[1]{\emph{\textcolor{b_skyblue}{#1}}}
\newcommand{\CSt}[1]{\emph{\textcolor{b_purple}{#1}}}
\newcommand{\Tonet}[1]{\emph{\textcolor{b_green}{#1}}}
\newcommand{\Modt}[1]{\emph{\textcolor{b_orange}{#1}}}
\newcommand{\Concernt}[1]{\emph{\textcolor{b_plum}{#1}}}

% Text underline
\newcommand{\TScode}[1]{{\textft{\customul[b_skyblue]{#1}}}}
\newcommand{\CScode}[1]{{\textft{\customul[b_purple]{#1}}}}
\newcommand{\Tonecode}[1]{{\textft{\customul[b_green]{#1}}}}
\newcommand{\Modcode}[1]{{\textft{\customul[b_orange]{#1}}}}
\newcommand{\Concerncode}[1]{{\textft{\customul[b_plum]{#1}}}}

\newcommand{\TVt}[1]{\emph{\textcolor{b_blue}{#1}}}
\newcommand{\TAt}[1]{\emph{\textcolor{b_red}{#1}}}
\newcommand{\TJt}[1]{\emph{\textcolor{b_green}{#1}}}

\newcommand{\TVcode}[1]{{\textft{\customul[b_blue]{#1}}}}
\newcommand{\TAcode}[1]{{\textft{\customul[b_red]{#1}}}}
\newcommand{\TJcode}[1]{{\textft{\customul[b_green]{#1}}}}

% Boxed and colored text (using \fbox for borders only)
\newcommand{\TS}[1]{\sethlcolor{b_skyblue!25!white}\hl{#1}}
\newcommand{\CS}[1]{\sethlcolor{b_purple!25!white}\hl{#1}}
\newcommand{\Tone}[1]{\sethlcolor{b_green!25!white}\hl{#1}}
\newcommand{\Mod}[1]{\sethlcolor{b_orange!25!white}\hl{#1}}
\newcommand{\Concern}[1]{\sethlcolor{b_plum!25!white}\hl{#1}}
\begin{abstract}
    Autistic individuals often experience negative self-talk (NST), leading to increased anxiety and depression. While therapy is recommended, it presents challenges for many autistic individuals. Meanwhile, a growing number are turning to large language models (LLMs) for mental health support. To understand how autistic individuals perceive AI's role in coping with NST, we surveyed 200 autistic adults and interviewed practitioners. We also analyzed LLM responses to participants' hypothetical prompts about their NST. Our findings show that participants view LLMs as useful for managing NST by identifying and reframing negative thoughts. Both participants and practitioners recognize AI's potential to support therapy and emotional expression. Participants also expressed concerns about LLMs' understanding of neurodivergent thought patterns, particularly due to the neurotypical bias of LLMs. Practitioners critiqued LLMs' responses as overly wordy, vague, and overwhelming. This study contributes to the growing research on AI-assisted mental health support, with specific insights for supporting the autistic community.
\end{abstract}




\maketitle

\section{Introduction}
Negative self-talk (NST) is the inner voice that often manifests as self-doubt, disappointment, or critical thoughts about oneself \cite{ronan1997self}. NST can significantly impact emotional and mental well-being, as it tends to be repetitive, intrusive, and can lead to rumination \cite{kendall1989self,calvete2002self,reichl2013relation}. This pattern of negative thinking is a significant predictor of mental health conditions, such as anxiety and depression \cite{michl2013rumination,nolen1991responses}. Prior research shows that autistic individuals\footnote{In this paper, we use identity-first language ('autistic individual') rather than person-first language ('individual with autism') to align with the preferences of some autistic individuals, while acknowledging other preferences \cite{kenny2016terms}.} are more likely to experience repetitive negative thinking compared to their non-autistic peers \cite{cooper2022cognitive,mckenney2023repetitive}. Given the higher rates of co-occurring conditions among autistic individuals \cite{brookman2018characterizing,crane2019something,camm2019people} compared to the general population, medical practitioners and clinicians working with autistic clients and members of the autistic community themselves often stress the importance of addressing NST to improve mental health. Although therapy is a recommended approach for coping with NST \cite{scarpa2013cbt,kendall2016mediators}, it may fall short for some autistic individuals, particularly those who struggle with verbal communication or have difficulty understanding or expressing emotions \cite{arnold2024barriers,mazurek2023mixed}. 

Given these limitations in traditional therapy methods, and the growing accessibility of large language models (LLMs), such as ChatGPT, Gemini, and Claude, many autistic individuals use these tools to help with interpersonal communication, such as explaining or interpreting social situations \cite{stokel-walker_ai_2023,harwell2022stumbling,jang2024s} and better understanding and processing their own emotions and thoughts \cite{Hoover}. Furthermore, an increasing number of autistic users rely on LLMs to discuss personal issues and seek mental health guidance \cite{carik2024exploring,mcfayden2024chatgpt,choi2024unlock}. However, both mental health professionals and members of the autistic community have raised concerns about the safety of relying on these tools for mental health support \cite{khawaja2023your,camm2019people}, particularly due to potential neurotypical biases in the LLM responses \cite{killian2023knock,weidinger_2022} and negative consequences observed in previous uses of chatbots in mental health care \cite{brusselstimes2024chatgpt,guardian2023chatbot}.

Despite these valid concerns, LLMs hold the potential to help autistic individuals express their thoughts and emotions more effectively, especially by incorporating modalities beyond traditional verbal communication \cite{arnold2024barriers,mazurek2023mixed}. This raises the question of whether AI can assist autistic individuals in coping with NST, potentially improving their therapeutic experiences without replacing mental health professionals. To explore this potential, it is essential to consider both autistic individuals' and mental health practitioners' perspectives, identifying key challenges in current LLM models from the therapist's viewpoint. With this in mind, we ask the following research questions:

\begin{itemize}
    \item RQ1. How do autistic individuals experience NST, and what coping strategies do they use?
    \item RQ2. How do autistic individuals envision the role of AI in coping with NST?
    \begin{itemize}
        \item[] RQ2.1. What are the interaction preferences of autistic individuals when using AI for coping with NST?
        \item[] RQ2.2. What concerns do autistic individuals have regarding the use of AI for coping with NST?
        \item[] RQ2.3. How do mental health practitioners view these perspectives from autistic individuals, and perceive AI’s potential to support their current therapeutic practices for their autistic clients?
    \end{itemize}
    \item RQ3. How do autistic individuals prompt AI about their NST, and how do therapists perceive the AI’s responses?
\end{itemize}

To address our research questions, we conducted two studies. First, we carried out an online survey of 200 autistic adults. This was followed by semi-structured interviews with three practitioners who have extensive experience working with autistic clients in their current roles. The practitioners included two therapists specializing in Occupational Therapy (OT), Cognitive Behavioral Therapy (CBT), and Acceptance and Commitment Therapy (ACT), as well as an expert in community-based support services. The survey explored participants' experiences with NST, their desired LLM support for coping with NST, preferences for LLM interaction in NST coping, and concerns about using LLMs for NST support. Participants also provided hypothetical prompts they would use to initiate a conversation with LLMs for NST support. We then input these prompts into ChatGPT and Claude to generate responses. In the interview study, we presented our survey findings and the LLM-generated responses to the practitioners. They evaluated the potential of LLM to support their autistic clients, its integration into their current therapeutic practices, and discussed issues with LLM outputs based on the participants' prompts.

Our study shows that autistic individuals frequently experience NST revolving around three primary themes: negative self-perception, perceived social judgment, and social disconnection (\textbf{RQ1}). Participants' NSTs are often triggered by work-related stress, disruption in their routine, or challenges in social interactions, and contribute to their anxiety, depression, demotivation, and social avoidance. To cope, participants adopt various strategies, with distraction techniques being the most common, followed by talking to friends or family, therapy, mindfulness, and positive self-talk. Many also turn to technology, using streaming services, meditation apps, and AI chatbots, though some struggle to find effective coping mechanisms.

Our findings also show that many of our autistic participants envision LLMs as a potential support mechanism for managing their NST (\textbf{RQ2}). They anticipate LLMs could assist in identifying their negative thought patterns, reframing these thoughts, guiding them through meditation and relaxation exercises, and serving as a conversational partner to help articulate their feelings, especially during periods of intrusive thoughts. Both autistic participants and practitioners see the value of LLMs in multiple contexts: as an additional support system during and between therapy sessions, as a tool to help verbalize emotions more clearly to therapists, and to facilitate the completion of take-home therapy practices. Many participants also expressed a strong interest in diverse interaction modalities beyond the current text-centric nature of commercial LLMs. Participants envisioned creative LLM support systems capable of providing spoken responses, generating music, offering tactile feedback, and creating visual illustrations in the context of coping with their NST. Participants' desire for multimodal interactions reflects a recognition that managing NST often requires multi-sensory approaches \cite{unwin2022use,dubois2017assessing}, aligning with the varied sensory preferences and processing styles common among autistic individuals \cite{dubois2017assessing}. Participants’ preferences for LLM conversation structure and tone also varied widely, reflecting the diverse spectrum of communication preferences and styles. Yet, many raised concerns around LLMs' ability to understand their neurodivergent thought patterns and feelings, particularly given the perceived neurotypical discursive styles in commercial LLMs. Despite these concerns, most remained open to using LLMs for coping with NST, provided the systems could adapt to their individual discursive preferences.

Our analysis of practitioners' feedback on LLMs' responses to participants' hypothetical prompts for coping with NST revealed three key categories: vague and abstract expressions, long list of choices and jargon-heavy language, and lack of actionable guidance (\textbf{RQ3}). Therapists and experts criticized the responses from both ChatGPT and Claude as overly wordy, vague, and potentially overwhelming. While they acknowledged that the suggestions were rooted in evidence-based practices, they emphasized the potential risks of applying these recommendations without professional guidance and professional support. 

In summary, we make the following contributions in this paper:

\begin{enumerate}
    \item An empirical understanding of autistic individuals' experience with NST and their coping strategies based on a survey of 200 autistic adults,
    \item An evaluation of how autistic adults envision AI support for coping with NST,
    \item Identification of autistic individuals' interaction preferences and concerns when interacting with LLMs for coping with NST,
    %\item Preliminary information on commercial LLMs' responses to participants' prompts from a practitioners' perspective,
    \item  Design implications for inclusive AI mental health support tool that complement rather than replace therapy, balancing between user desire for personalization and their preferences for safety and privacy.
\end{enumerate}
\section{Related Work}

\subsection{NST and Autism}
Autism affects approximately 1 in 36 youth in the U.S. and is recognized as a lifelong neurodevelopmental condition characterized by differences in social interactions, along with restricted and repetitive behaviors, interests, or activities \cite{maenner2023prevalence}. Research has highlighted how these traits may contribute to higher rates of repetitive thinking among autistic individuals \cite{zappullo2023role} with studies showing that autistic people tend to experience more negative repetitive thinking compared to their non-autistic counterparts \cite{cooper2022cognitive,mckenney2023repetitive}. Such NST and rumination are known to be strong predictors of anxiety, depression, and physiological distress \cite{michl2013rumination,nolen1991responses}. Given that an estimated 90\% of autistic individuals in clinical samples meet the criteria for at least one additional mental health condition, including anxiety and depression \cite{brookman2018characterizing}, addressing this issue is a significant priority for this community.
 
Therapeutic approaches, like CBT and ACT, are commonly recommended for coping with NST. For example, CBT focuses on helping individuals reframe negative thoughts to develop more constructive and positive ways of thinking \cite{scarpa2013cbt,kendall2016mediators}. However, traditional therapy may not always be effective for autistic individuals. Verbal interactions, a core element of therapies like CBT and ACT, can present challenges for many autistic people, who may struggle with understanding and expressing their thoughts, feelings, and intentions, as well as those of others \cite{cummins2020autistic,ekman2015modified}. This difficulty extends to articulating their emotional experiences and thoughts, which can hinder the therapeutic process. A recent study also suggests that many therapeutic techniques can be challenging for autistic individuals to apply in daily life \cite{mazurek2023mixed}. Moreover, access to therapy may be limited due to factors such as cost and availability \cite{singla2023democratizing,fonagy2021socioeconomic}. In light of these challenges, this study explores how AI could assist autistic individuals in coping with NST, investigating its potential role in supporting both mental health and therapy processes for autistic individuals and therapists.

\subsection{HCI Research in Mental Health and Autism}
The HCI community has explored incorporating technologies to improve therapeutic outcomes for autistic individuals, including information and communications tools \cite{scarcella2023information}, such as games \cite{malinverni2017inclusive,ahmed2023applications,tang2019systematic}, social robots \cite{dipietro2019computer,kouroupa2022use,sartorato2017improving,salimi2021social}, or chatbots \cite{lee2019caring,fitzpatrick2017delivering}, virtual and augmented reality (VR/AR) \cite{krishnappa2022gaze,falconer2016embodying,mesa2018effectiveness,karami2021effectiveness,berenguer2020exploring,almurashi2022augmented}, wearables \cite{washington2017superpowerglass,koumpouros2019wearables,taj2018review}, and tactile strategies \cite{miri2022far,croy2016affective}. For instance, games like StarRescue \cite{bei2024starrescue} aim to improve turn-taking and social communication skills among autistic children through collaborative gameplay, while RoutineAid \cite{kim2024narrating} gamifies daily tasks to support healthy routines and promote their emotional and social well-being in autistic adults. Similarly, VR has been shown to offer controlled environments that reduce overwhelming sensory input, making interactions more comfortable compared to traditional therapy approaches \cite{boyd2018vrsocial}. Wearables have also demonstrated promise in improving emotional recognition and social engagement in home-based therapy \cite{washington2017superpowerglass}. 

Recent advancements in AI have further contributed to these therapeutic interventions \cite{rehman2021features,tang2024emoeden}, particularly through the integration of social robots in CBT \cite{marino2020outcomes} and the incorporation of music and art therapies, which are reported to be effective approaches for autistic individuals \cite{martinez2024interventions,johnston2018innovative,ragone2020designing,bauer2024musictraces}. Integration of AI-driven features, such as music generation, tone transfer, and transforming sketches into realistic images or artistic styles, expand the potential of these therapeutic methods \cite{sun2024understanding,du2024deepthink}. Despite these advancements, much of the existing research has primarily focused on autistic children, with an emphasis on diagnosis, communication, and social skill development \cite{howlin2015research,howlin2015addressing,lorenc2018support}. 

Recent studies, however, have started to examine the challenges faced by autistic adults across broader domains and how technology can provide meaningful support. For instance, Ara et al. \cite{ara2024collaborative} described how collaborative, communication-oriented tools can reduce barriers encountered during the job-seeking process. Similarly, Cullen et al. \cite{cullen2024towards} studied how autistic young adults regulate their social media use, particularly the dynamics between self-regulation strategies and the involvement of their support networks. Similarly, Williams et al. \cite{williams2023cyborg} explored how autistic adults integrate various technologies and social supports to manage cognitive and executive function challenges in their daily lives. Zolyomi et al. \cite{zolyomi2024emotion} studied emotion dyads to inform the design of tools that support emotional understanding and effective communication in neurodiverse interactions.

Despite this growing body of research, there has been relatively less attention given to the mental health needs of autistic adults, many of whom are increasingly turning to AI and LLM tools for companionship and mental health support \cite{mcfayden2024chatgpt,choi2024unlock}. On online platforms like Reddit and X (formerly Twitter), autistic individuals frequently share their experiences with these tools \cite{jang2024s,egner2022actuallyautistic,jaiswal2024using}, highlighting their availability, non-judgmental nature and clarity in responses \cite{jang2024s,martin2024bridging}. However, concerns have also been raised, particularly around safety \cite{de2024chatbots,miner2019key}, neurotypical bias \cite{killian2023knock,weidinger_2022}, and the lack of personalized responses \cite{nazir2023comprehensive}. As discussions around these tools continue to grow, this study aims to point out the potential challenges posed by commercial LLM responses from therapists’ perspectives. We further explore, through a survey of autistic individuals, how they envision AI tools supporting their mental health, especially in coping with NST.
\section{Methodology}
In this work, we conducted an online survey with autistic adults to explore their experiences with NST and their perspectives on how AI could support them cope with these thoughts. The survey further delved into the types of support mechanisms they preferred, their interaction preferences, and concerns they have about using AI for coping with NST. Building on the survey insights, we consulted with practitioners who work closely with autistic individuals. These consultations aimed to assess participants' answers from a clinical standpoint and explore the potential integration of AI into their current therapeutic practices. During these discussions, we also shared with practitioners the participants' prompts and responses from two widely used commercial AI tools, ChatGPT \cite{openai2024chatgpt} and Claude \cite{anthropic2024claude}, to point out the problems with these responses.

\subsection{Survey Study}
\subsubsection{Survey Design}
Our survey consists of six sections designed to explore participants' experiences with NST and their perception of AI supporting them in coping with NST. The complete list of survey questions is provided in the Appendix, Tables \ref{tab:ATQ_N_items} and \ref{tab:survey_questions}. To ensure that the survey questions aligned with participants’ experiences and prioritized their safety, we collaborated with two clinical psychologists on the author team to select the validated instruments and co-design the questions. We also sought feedback from a self-advocate autistic advisory committee to refine the survey, ensuring it is neuro-affirming and representative of autistic individuals’ diverse experiences.

The first section of the survey asked participants about their general perceptions of AI, focusing on their views of AI as a tool for mental health support. We assessed their comfort level with AI and how effective they believe AI could be in this role (5 questions). 

In the second section, participants answered two psychological questionnaires to measure the frequency of NST: Negative Automatic Thought Questionnaire (ATQ-N) and Anxiety Scale for Autism-Adults (ASA-A). The ATQ-N is a widely used and validated instrument in psychological research for assessing the frequency of negative thoughts \cite{netemeyer2002psychometric}. For this study, we selected the 8-item version (ATQ-8) to minimize potential emotional distress while maintaining strong psychometric properties, as demonstrated by its internal consistency (Cronbach’s $\alpha = 0.89$) \cite{ruiz2020psychometric}. This short version has also been validated for autistic populations, showing high internal consistency (Cronbach’s $\alpha = 0.97$) \cite{yang2023pilot,baires2023cultural}.

The ASA-A is another validated instrument specifically designed with input from autistic adults to capture their unique experiences of anxiety \cite{rodgers2020development,cassidy2021development}. From this scale, we selected the Uncertainty and Social Anxiety factors (11 items), as these cognitive patterns are closely linked to negative thoughts in prior research \cite{nishikawa2022social,mcevoy2013intolerance}. The ASA-A demonstrated strong internal consistency (Cronbach’s $\alpha = 0.93$), with excellent reliability for these factors (Cronbach’s $\alpha = 0.92$ for Social Anxiety and $\alpha = 0.90$ for Uncertainty) \cite{rodgers2020development}. Together, these two scales resulted in 19 NST items, rated by participants for their frequency over the past week.

Due to the highly personal nature of NST \cite{lefebvre2022interpersonal}, participants were given the option to provide their own self-written NST if as an open-text response to account for any potential NST missing from the 19 items, and to indicate the frequency of its occurrence in the past week. Out of the most frequently occurring NST items, including their own, participants were then asked to select the one that affected them the most. Next, the survey asked participants to provide further details about what triggers this particular NST, how it affects their daily life (including relationships and work or school performance), and what strategies they use to cope when this thought occurs (10 questions). 

In the third section, we threaded their most challenging NST chosen in the second section to ask participants about their preferences for how they envision LLMs assisting them in coping with this particular NST and their desired LLM interaction preferences, including conversation structure, tone, and modality (9 questions). We provided structured items to rate along with open-ended questions to better elaborate on their preferences and needs.

In the fourth section, participants were asked how they would start a conversation about this particular NST with an LLM. They were requested to provide a prompt as open-text responses, rate the difficulty of writing the prompt, and share what made it difficult for them to do so. 

The final two sections addressed participants' concerns about using LLMs for coping with NST (3 questions) and their familiarity with AI (4 questions). Across the survey, we used a 5-point scale for all non-open-ended questions.

\begin{table}[ht]
    \centering
        \caption{Demographics of study participants (N=200).}
    \begin{tabular}{p{5cm}|c}
         \textbf{Demographic Category} & \textbf{\textit{N} (\% of sample)} \\
         \toprule
         
         \textbf{Race/Ethnicity} & \\

            \hspace{1em} White/Caucasian & 141 (70.5\%) \\
            \hspace{1em} Multiracial & 24 (12\%) \\
            \hspace{1em} Black/African American & 17 (8.5\%) \\
            \hspace{1em} Hispanic or Latino & 8 (4.0\%) \\
            \hspace{1em} Asian & 5 (2.5\%) \\
         \midrule
         
         \textbf{Gender} & \\
            \hspace{1em} Male & 86 (43.0\%) \\
            \hspace{1em} Female & 81 (40.5\%) \\
            \hspace{1em} Multigender & 9 (4.5\%) \\
            \hspace{1em} Nonbinary & 6 (3.0\%) \\
            \hspace{1em} Agender & 5 (2.5\%) \\
            \hspace{1em} Genderqueer & 4 (2.0\%) \\
            \hspace{1em} Transgender & 4 (2.0\%) \\
            \hspace{1em} Other & 4 (2.0\%) \\
         \midrule
         
         \textbf{Age} & M = 37.22,\\
         & SD = 11.90 \\
         \midrule

         \textbf{Education} & \\
            \hspace{1em} Bachelor's degree & 56 (28.0\%) \\
            \hspace{1em} Some college & 42 (21.0\%) \\
            \hspace{1em} Master's degree & 29 (14.5\%) \\
            \hspace{1em} High school graduate & 26 (13.0\%) \\
            \hspace{1em} Associate degree & 19 (9.5\%) \\
            \hspace{1em} Some postgraduate work & 14 (7.0\%) \\
            \hspace{1em} Less than high school & 6 (3.0\%) \\
            \hspace{1em} Doctorate & 4 (2.0\%) \\
            \hspace{1em} Other & 4 (2.0\%) \\
         \midrule
         
         \textbf{Employment} & \\
            \hspace{1em} Full-time employment & 67 (33.5\%) \\
            \hspace{1em} Part-time employment & 42 (21.0\%) \\
            \hspace{1em} Unable to work & 33 (16.5\%) \\
            \hspace{1em} Unemployed and currently looking for work & 33 (16.5\%) \\
            \hspace{1em} Unemployed and not currently looking for work & 19 (9.5\%) \\
            \hspace{1em} Retired & 6 (3.3\%) \\
         \midrule

         \textbf{Income} & \\
            \hspace{1em} < \$10,000 - \$39,999 & 101 (50.5\%) \\
            \hspace{1em} \$40,000 - \$69,999 & 45 (22.5\%) \\
            \hspace{1em} \$70,000 - \$99,999 & 27 (13.5\%) \\
            \hspace{1em} \$100,000 - \$150,000> & 27 (13.5\%) \\
         \bottomrule
    \end{tabular}
    \label{tab:demographics}
    \Description{This table summarizes the demographic characteristics of the study participants (N=200). The majority of participants identified as White/Caucasian (70.5\%), with a smaller percentage identifying as multiracial (12\%) or other races. The gender distribution was 43\% male and 40.5\% female, with representation from nonbinary, multigender, and other gender identities. The average age was 37.22 years (SD = 11.90). Most participants had a bachelor’s degree (28\%) and were in full-time employment (33.5\%). Income levels varied, with 50.5\% earning less than \$40,000 annually.}
\end{table}

\subsubsection{Participants Recruitment} 
We recruited 200 participants who met the following criteria: (1) identify as autistic or have a formal autism diagnosis, (2) are 18 years of age or older, (3) are comfortable reading, writing, and participating in the survey in English, (4) are able to provide informed consent independently, and (5) have experience with NST. 

We did not require participants to have prior LLM experience to capture broad perspectives, including those who might benefit from LLM-driven support but have not yet engaged with such tools. Non-users may have different expectations, concerns, and perceptions that are not influenced by prior experience, which may reveal unmet needs or design opportunities.

Recruitment was conducted through SPARK (Simon Powering Autism Research), the largest autism study platform in the U.S. that connects researchers with autistic individuals and their caregivers \cite{SPARK_Autism}. Recruitment flyers outlined the study's purpose as learning about how autistic individuals experience NST and their perception towards the support of AI technology to cope with NST. We informed all participants that their responses might be used for analysis. The survey took approximately 40 minutes to complete, and each participant received a \$10 gift card as compensation for their time.

\subsubsection{Participants Demographics}
Participants were recruited from the U.S. with the majority identifying as white or Caucasian (70.4\%). The gender distribution was relatively balanced, with 43\% and female 40.5\%. The participants were aged between 18 to 80, with the majority in the 20-40 range (mean = 37.22, SD = 11.90). The education level among participants was high, with about 70\% holding a bachelor's degree or higher. Despite this, only 54.5\% reported being employed full-time or part-time, and 73\% of them have an annual income of less than \$70,000. 

Almost all participants (94\%) reported having at least one co-occurring condition, with over 60\% reporting three or more. The most common diagnoses were generalized anxiety disorder, attention-deficit/hyperactivity disorder (ADHD), major depressive disorder, and post-traumatic stress disorder (PTSD).

Nearly all of our participants are familiar with AI, and among them, 57\% reported using AI at least once a week, primarily for seeking information, entertainment, and communication with others. Additionally, 17\% of the participants indicated that they use AI for mental health support and emotional check-ins. Of this group, a significant majority (74\%) find AI to be useful for these tasks.

\begin{table}[ht]
    \centering
    \small
    \caption{Demographic and professional information of practitioners, including their roles, areas of practice, years of experience, and gender.}
    \begin{tabular}{p{0.1\textwidth} p{0.2\textwidth} >{\centering}p{0.045\textwidth } p{0.045\textwidth} c}
        \toprule
        \textbf{Role} & \textbf{Area of Practice} & \textbf{Years} & \textbf{Gender} \\
        \midrule
        OT Practitioner & Occupational Therapist (OT) & 20 & Female \\
        \midrule
        ACT/CBT \newline Practitioner & Acceptance and Commitment Therapy (ACT) \& \newline Cognitive Behavioral Therapy (CBT) & 10 & Female \\ 
        \midrule
        CSS \newline Practitioner & Community-Based Support \newline Services (CSS) & 20 & Female \\
        \bottomrule
    \end{tabular}
    \label{practitioner_demographic}
    \Description{Table 2 provides an overview of the demographic and professional backgrounds of the practitioners involved in the study. The practitioners include an Occupational Therapist (OT) with 20 years of experience, a practitioner specializing in Acceptance and Commitment Therapy (ACT) and Cognitive Behavioral Therapy (CBT) with 10 years of experience, and a practitioner focusing on Community-Based Support Services (CSS) with 20 years of experience. All practitioners identify as female. }
\end{table}

\subsection{Practitioner Evaluation}
After completion of the survey, we discussed our results with practitioners who work closely with autistic individuals to gain their perspectives on the impact of NST on their clients' progress and their views on the role of AI in supporting autistic individuals in managing NST. The practitioners included an Occupational Therapist (OT) with over 20 years of experience supporting autistic individuals through sensory integrative approaches, a clinical researcher with more than 10 years of experience specializing in Cognitive Behavioral Therapy (CBT) and Acceptance and Commitment Therapy (ACT), and a researcher and administrator with over 20 years of experience in community-based support services (CSS) and intervention development for promoting independent living among autistic individuals. The demographic and professional details of these practitioners are summarized in Table \ref{practitioner_demographic}.

The selection of these therapeutic approaches was guided by the two clinical psychologists on our author team, who prioritized methods known for their effectiveness in addressing NST and their widespread use in autism support.

First, we conducted a group discussion where practitioners collectively reviewed the survey results, including participants' experiences with NST, their AI-related preferences, and concerns. This phase established a shared understanding of the findings.

In the second phase, we conducted individual sessions, each lasting over an hour, where practitioners provided an in-depth assessment of the open responses of the participants about the types of support and the different modalities they envisioned AI providing to cope with NST. They evaluated the potential benefits of these methods for autistic individuals, how they could be integrated into their current therapeutic practices, and the potential impact on their clients. 

As part of these sessions, we examined how practitioners interpreted and evaluated the ways participants initiated discussions about NST with LLMs. To facilitate this, we first analyzed all 128 prompts provided in the survey and identified recurring patterns, such as directly sharing NST, asking for specific strategies, or requesting direct help. Based on this analysis, we selected representative prompts that reflected both common and varied ways participants framed their NST-related concerns. The categories, example prompts, and LLM responses are provided in Appendix Tables \ref{tab:llm_comparison} and \ref{tab:llm_comparison_2}.

Before showing the prompts, we asked practitioners how they would address these thoughts if raised by a client during therapy. We then shared LLM-generated responses from ChatGPT and Claude as both autistic individuals and researchers use these tools \cite{Hoover,mcfayden2024chatgpt,bertacchini2023social}. Although previous research showed the neurotypical biases in these responses, many neurodivergent individuals, including those in the autistic community, turn to LLMs for mental health support \cite{Hoover,carik2024exploring}. We asked practitioners to identify any issues or potential risks in these LLM-generated responses to understand these concerns from medical practitioners' perspectives.

\subsection{Data Analysis}
We analyzed both quantitative and open-text survey responses. For quantitative questions, we calculated percentages based on the number of participants selecting each option and reported mean values. For the open-ended survey responses, we applied an inductive coding approach, where two researchers independently developed and refined codes based on participants' qualitative input. The research team then reviewed and finalized the themes, ensuring that responses complemented the structured survey data by allowing participants to elaborate on their visions regarding AI-related questions.

To analyze practitioner feedback, the first author transcribed the meetings, and two researchers independently coded the transcripts to identify emerging themes using an inductive thematic analysis approach. They refined the codes through iterative discussions, collaboratively resolving discrepancies to ensure an accurate representation of practitioners’ insights. The research team then reviewed and consolidated the final themes, ensuring they captured key concerns, potential risks, and therapeutic considerations discussed by practitioners.

\subsection{Ethical Considerations}
This study was approved by Virginia Tech's Institutional Review Board (IRB). To ensure participants' safety and well-being, we also collaborated with a self-advocate autistic advisory committee (SAAC) throughout the research process. We sought feedback from the SAAC to align our research objectives and survey questions with the priorities of autistic individuals, making the study neuro-affirming, inclusive, and meaningful. The committee was co-headed by one of the authors, a therapist with more than 30 years of experience working with autistic individuals, specializing in CBT and ACT. 

We also recognize that reflecting on negative thoughts and experiences may cause stress or discomfort for participants. To address this, we provided a comprehensive list of resources, including national and local support services such as the Suicide and Crisis Lifeline, which were accessible to participants before, during, and after the study. Participants were also informed at the start of the study that they could withdraw or skip any questions at any time. Additionally, one of the authors, a licensed clinical psychologist, monitored the entire process to address any concerns that arose as a result of the study.
\begin{figure*}[!ht]
    \centering
    \includegraphics[width=0.9\textwidth]{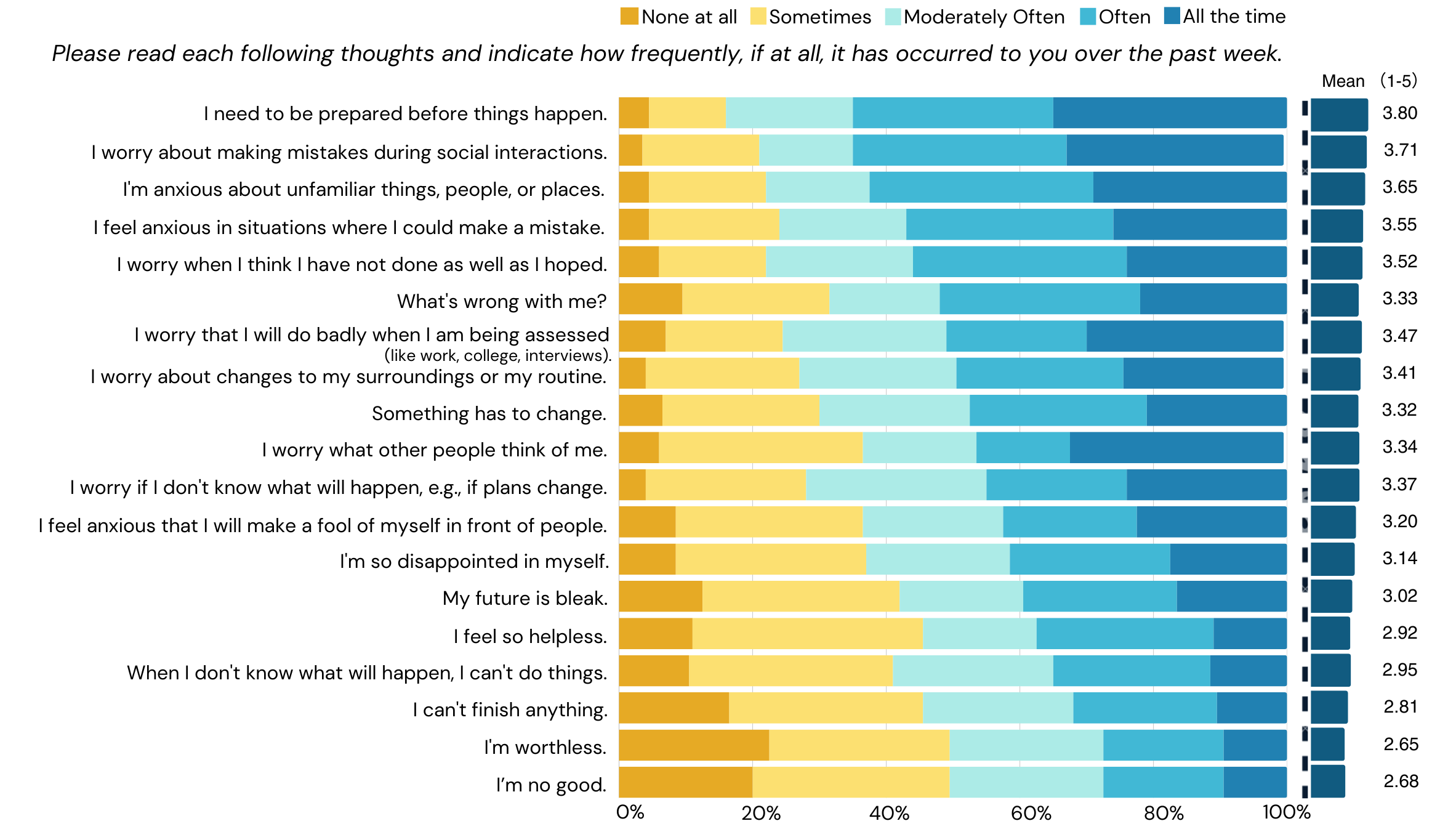}
    \caption{Frequency of NST based on responses from the Negative Automatic Thoughts (ATQ-N 10) and Anxiety Scale for Autism-Adults (ASA-A). Distribution of participants' responses to the items in these questionnaires, with an overall mean frequency score of 3.25 (on a scale of 1–5). Participants reported a high frequency of thoughts related to the need for preparation, anxiety about social interactions, and discomfort with unfamiliar situations.} 
    \label{fig:NST_frequency}
    \Description{This figure presents a stacked bar chart showing the frequency of NST reported by participants. The chart includes multiple statements, each represented by a horizontal bar. The bars are color-coded to indicate the frequency categories: "None at all" (light yellow), "Sometimes" (light orange), "Moderately often" (blue), "Often" (dark blue), and "All the time" (deep blue). The percentage of responses in each frequency category is displayed along the horizontal axis, ranging from 0\% to 100\%. Each statement's mean score is shown on a 1–5 scale to the right of the bars. The statements cover themes such as needing to prepare before things happen, worrying about making mistakes in social interactions, feeling anxious about unfamiliar places or situations, and self-critical thoughts like "I'm worthless" or "I'm no good." Statements associated with preparation and social anxiety have the highest mean scores (e.g., "I need to be prepared before things happen" with a mean of 3.80), while statements like "I'm no good" have the lowest mean scores (2.68). The chart caption notes an overall mean frequency score of 3.25 (on a scale of 1–5), highlighting participants' high frequency of thoughts related to the need for preparation, anxiety about social interactions, and discomfort with unfamiliar situations.}
\end{figure*}
\section{Findings}

To answer our research questions, we analyzed both quantitative and qualitative data from the survey and discussions with practitioners. We first presented the most common NST experienced by participants, focusing on their triggers, the impact on their daily lives, and the coping strategies they employed (RQ1). We then explored how autistic individuals envision AI's role in coping with NST, including their preferences for conversation structure, tone, and modalities, followed by mental health practitioners' evaluations of these visions (RQ2). Finally, we showed participants' prompts to the LLMs and their responses, along with evaluations from practitioners (RQ3). 

\subsection{Participants' Experience with NST (RQ1)}
The results of the questionnaires reveal that NST is a significant issue among autistic individuals, as participants reported experiencing these thoughts more than moderately often on average. The overall mean frequency score across all NST items was 3.25 (on a scale of 1–5, where 1 = None at all and 5 = All the time), indicating a high prevalence of these thoughts in participants' daily lives. Figure \ref{fig:NST_frequency} provides a detailed breakdown of how frequently participants experience specific NST, based on the items from the ATQ-N \cite{netemeyer2002psychometric} and the ASA-A \cite{rodgers2020development} scales. The most pervasive NST among participants is the need for preparation (\textit{``I need to be prepared before things happen''}), anxiety about social interactions (\textit{``I worry about making mistakes during social interactions''}), and discomfort with unfamiliarity (\textit{``I'm anxious about unfamiliar things, people, or places''}). Additionally, many participants frequently worried about how others perceive them (\textit{``I worry what other people think of me''}) and feared under-performance in work or academic settings (\textit{``I worry that I will do badly when I am being assessed, like work, college, or interviews''}). Given the personal nature of NST, 60 participants also provided their personal negative thoughts (not included in the 19 NST items) as open-text responses, which were qualitatively analyzed, resulting in three primary themes: negative self-perception, perceived social judgment, and social disconnection.

\begin{figure*}[ht]
    \centering
    \includegraphics[width=0.9\textwidth]{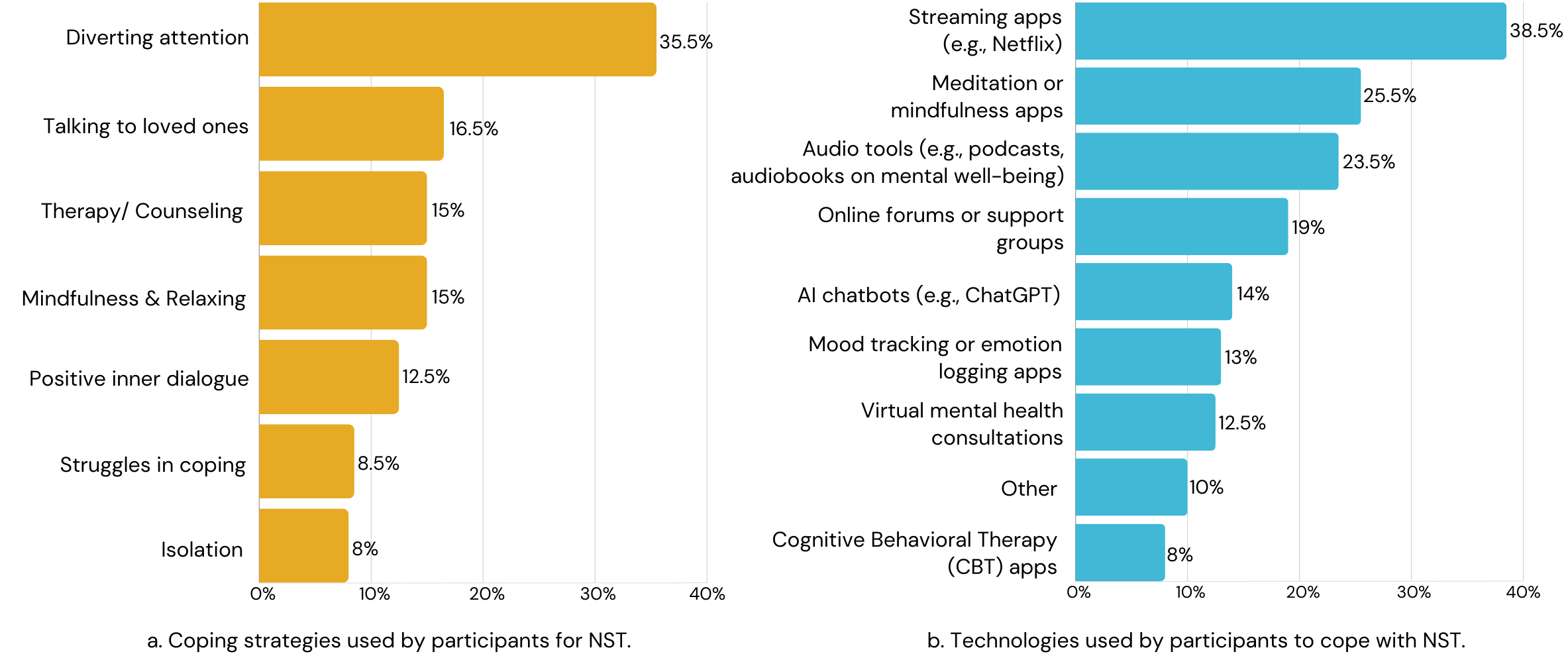}%{figures/tech_for_coping.png}
    \caption{Coping strategies and technologies used by participants to manage NST. Figure a (left) summarizes coping strategies derived from participants’ open-ended responses, where similar answers were grouped into broader categories. Figure b (right) shows the range of technologies participants reported using as coping mechanisms for NST, with an average of 2.3 tools selected by participants.}
    \label{fig:coping_mechanisms}
    \Description{This figure consists of two bar charts. The first chart (Figure 2a) displays coping strategies for negative self-talk (NST) derived from participants' open-ended responses. Strategies were grouped into categories, with "Diverting attention" being the most frequently reported (35.5\%), followed by "Talking to loved ones" (16.5\%), "Therapy/Counseling" (15\%), and "Mindfulness & Relaxing" (15\%). The second chart (Figure 2b) shows technologies participants use to cope with NST, based on a multiple-selection question. Streaming apps (e.g., Netflix) were the most commonly selected (38.5\%), followed by meditation or mindfulness apps (25.5\%), audio tools like podcasts or audiobooks (23.5\%), and online forums or support groups (19\%). AI chatbots (14\%) and mood-tracking apps (13\%) were also mentioned, while virtual mental health consultations (12.5\%) and CBT apps (8\%) were less frequently used.}
\end{figure*}

\textbf{{Negative self-perception.}} Many participants described their NST as revolving around feelings of inadequacy. For example, P121 shared their frustration about \textit{``having average grades despite their best efforts''}, which led them to think, \textit{``I will never be good at anything.''} Similarly, P32 expressed a constant fear of losing their job due to perceived underperformance, which triggered the recurring thought, \textit{``I just need to survive; I will not make it on my own.''}

Some participants reflected self-disappointment, particularly struggles with time management, prioritization, and focus, contributing to the feeling of \textit{``failing to achieve their goals''} (P7). For some, such as P76, this NST escalated into constant worry about their future success, articulated as, \textit{``I am constantly worried about providing myself with a secure future and fear that I cannot keep working full time''}.

Furthermore, many participants expressed feelings of unworthiness with thoughts like \textit{``I am less than human''} (P104) or \textit{``Why am I like this?''} (P89), often triggered by external stressors, such as hearing negative comments from others. These recurring thoughts contributed to the feeling of hopelessness, as captured by P72, stating, \textit{``Things will never change. I will always struggle. Why cannot I be normal?''} 

\textbf{Perceived social judgment.} Another recurring theme in the responses is the perception of negative social judgment. Participants frequently expressed thoughts, such as \textit{``Why does everyone hate me?''} (P12) or \textit{``The world is against me''} (P179). For those searching for a job, concerns about making a negative first impression during the job interview fueled these thoughts, with P12 explaining their worry that interviewers would \textit{``dislike them upon seeing or speaking to them.''} For some participants, simply leaving the house provoked intrusive thoughts like, \textit{``People are talking about me and making fun of me,''} (P36) or \textit{``People think I am stupid or slow''} (P194). These thoughts, for some participants, stemmed from the belief that others would notice something inherently wrong with them. For participants, like P54 and P87, such beliefs intensified the feeling of being a burden to those around them. Others like P69, these thoughts led them to \textit{``end up masking even more.''}

\textbf{Social disconnection.} Several participants expressed their challenges with communication, often caused by misunderstandings. P71 captured this sentiment, stating, \textit{``I do not understand why people do not understand me and why I do not understand them.''} This challenge often made it difficult for them to form and maintain relationships, as participants noted the stress of meeting new people or arguments with their significant other. For instance, P195 expressed persistent thoughts like \textit{``Why cannot I make friends or lasting relationships?''} while P82 described recurring \textit{``feelings of being left out''}.

Participants reported that all of these negative thoughts affected their relationships with family and partners, performance at work and school by decreasing motivation and productivity, and overall emotional well-being by leading to feelings of hopelessness and further anxiety and depression. Cumulative effects of these thoughts were summarized by P126 as \textit{``isolation, demotivated, no tasks done, house a mess, fail at school, not finding work, canceling appointments''}.

\begin{figure}[!ht]
    \centering
    \includegraphics[width=1.0\columnwidth]{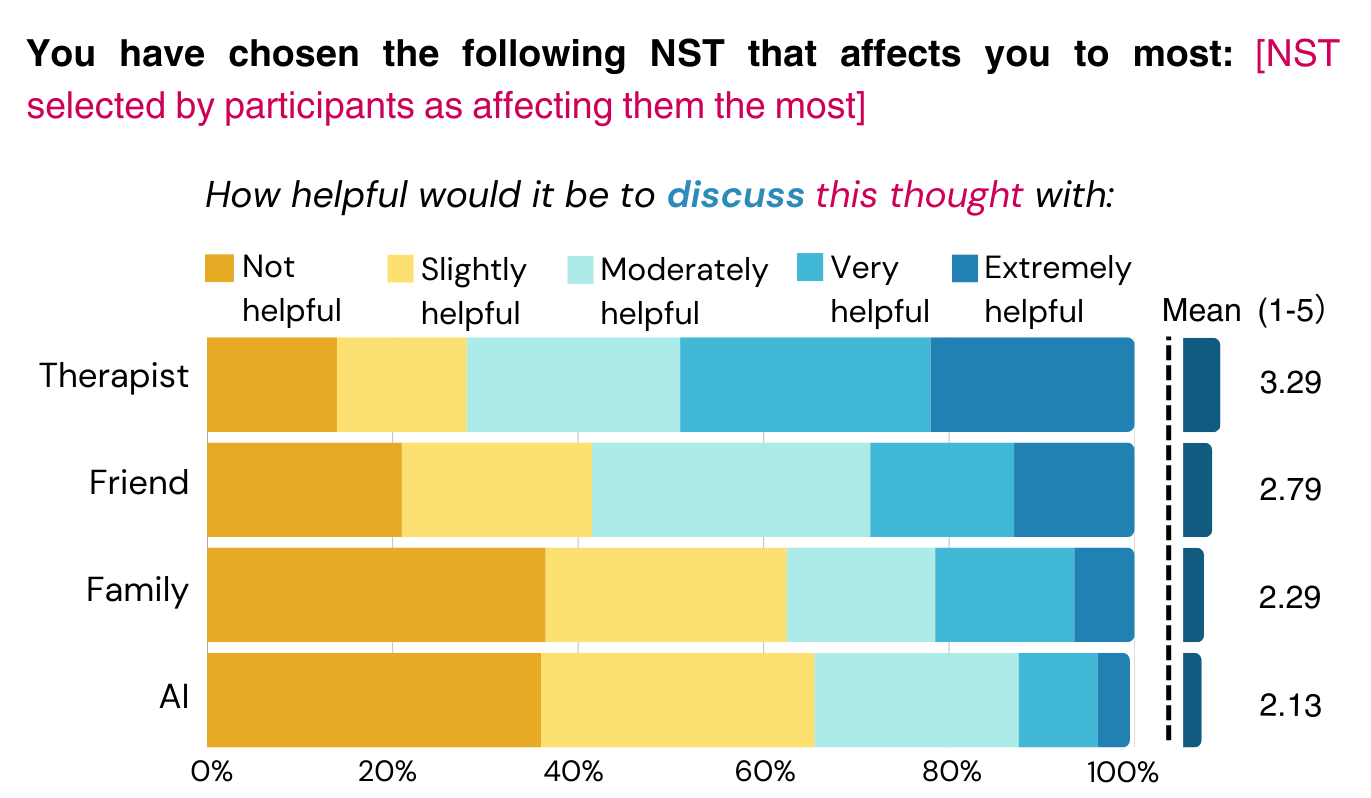}
    \caption{Participants rated the helpfulness of discussing their most impactful NST with different sources: a therapist, a friend or significant other, family, and AI, using a scale from 'Not helpful' to 'Extremely helpful.' Therapists received the highest mean helpfulness score (3.29), followed by friends or significant others (2.79), family (2.29), and AI (2.13). }
    \label{fig:talking_to}
    \Description{The figure presents a stacked bar chart comparing the perceived helpfulness of discussing NST with various sources: a therapist, a friend or significant other, family, and AI. Helpfulness was rated on a scale of five levels: Not helpful, Slightly helpful, Moderately helpful, Very helpful, and Extremely helpful. Each bar represents the distribution of responses for a specific source, with the segments proportionally scaled according to the number of responses in each category. For example, therapists received the highest proportions of "Very helpful" (27\%) and "Extremely helpful" (22\%), reflecting their overall high helpfulness ratings. In contrast, family and AI had the highest proportions of "Not helpful" responses, at 36.5\% and 36\%, respectively.}
\end{figure}

\subsubsection{Coping Mechanisms}
Participants reported various coping mechanisms for managing these negative thoughts in response to an open-ended question. Their responses were analyzed and grouped into broader categories to identify recurring themes. Participants also reported the technologies they use to support these efforts. On average, participants selected 2.3 different technologies for coping with NST, reflecting a tendency to rely on multiple tools for support. The breakdown of coping strategies and technologies is presented in Figure \ref{fig:coping_mechanisms}.

The most common strategy, reported by over 35\% of participants, is diverting attention through personal interests or hobbies. Many participants also use technology for distraction, such as watching content on streaming platforms or listening to music or podcasts. Some even get assistance from AI devices, asking, for example, \textit{``Alexa to play their mental health music Playlist''} (P23). 

However, the potential drawbacks of relying too much on distractions were highlighted by P146, who shared: \textit{``When I spend all day watching something or using chatbots, it distracts me from the impending doom. But in excess,  I stop caring about my health and responsibilities, which only makes things worse.''} The OT practitioner echoed this concern, noting that while streaming apps can provide relief, they may also intensify negative thoughts, as self-regulation can be particularly difficult for autistic individuals:

\begin{quote}
    OT Practitioner: \textit{``Before you know it, they have not eaten dinner, and it is 1 a.m., and they are thinking, 'Where did the time go?' Then they feel bad about the time they just wasted when it was supposed to be a mental health break or a distraction. This can lead to even more negative self-talk—internalizing thoughts like, 'I just wasted 6 hours when it was supposed to be a 30-minute break.'''}
\end{quote}

The second most common strategy, reported by 16.5\% of participants, is talking to their loved ones, such as family or friends. However, despite more than 60\% of participants expressing a desire to share their negative thoughts, 80\% reported having no one to talk to. Interestingly, unlike the general population, where most prefer discussing mental health with a close friend (57\%) \cite{pew2024mentalhealth}, our participants found talking with a therapist to be the most helpful compared to family or friends, as shown in Figure \ref{fig:talking_to}. Furthermore, 14\% of participants mentioned using AI chatbots like ChatGPT to talk or distract themselves from negative thoughts.

Meanwhile, around 15\% of participants attend therapy or counseling sessions to cope with these thoughts. Many also turned to apps for therapeutic practices, such as tracking mood and logging emotions, 12.5\% seeking virtual mental health consultancy, or 8\% practicing CBT techniques. Mindfulness and relaxation techniques are also common coping strategies among participants with 15\%, with over 25.5\% of participants using apps to support these practices. 

Some participants (12.5\%) practice positive self-talk by reassuring or encouraging themselves with phrases such as \textit{``You will be fine''} (P120), \textit{``I am doing well and that I know I will be able to do my best in these assignments''} (P94), or \textit{``Let them think what they want to think''} (P198). However, others noted difficulties finding or maintaining these positive thoughts. For instance, P127 stated, \textit{``it is hard to find [these positive thoughts] when feeling so negative''}, while P38 shared frustration that \textit{``despite fighting with those thoughts by reminding themselves of their worth, the thoughts never stop''.} Furthermore, several participants reported \textit{``not having coping mechanisms that actually work''} (P28) or \textit{``not sure about how to cope with these thoughts''} (P58).

\subsection{Participants' Visions for AI and Practitioners' View for These Visions (RQ2)}
\begin{figure*}[ht]
    \centering
    \includegraphics[width=1.0\textwidth]{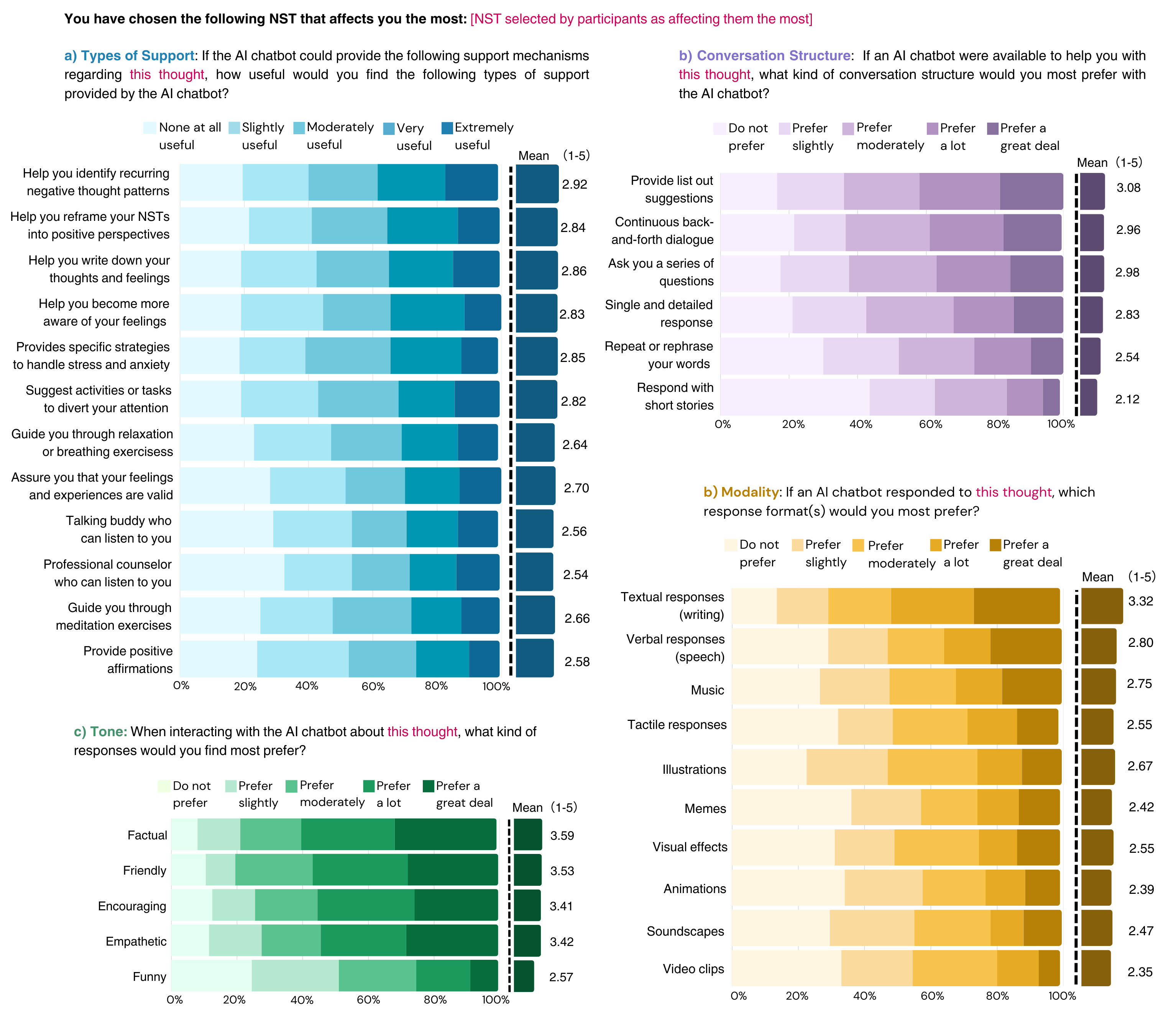}
    \caption{Participants' preferences for support types, conversation structures, tone, and modalities in LLM interactions for coping with NST. Mean ratings (1 = not at all useful, 5 = extremely useful) are displayed next to each bar to provide an overview of participants' evaluations for each category.}
    \label{fig:interac_pref}
    \Description{This figure illustrates participants' preferences for various aspects of AI chatbot support in coping with NST. The preferences are categorized into four panels: types of support, conversation structures, interaction modalities, and tone. In the Types of Support panel (top-left), participants rated the usefulness of different mechanisms, with "Helping to identify recurring negative thought patterns" receiving the highest mean score (2.92 on a scale of 1–5), while options like "Providing positive affirmations" and "Guiding through meditation exercises" were rated lower. The Conversation Structure panel (top-right) highlights preferences for styles of interaction, with "Providing a list of suggestions" (mean = 3.08) and "Continuous back-and-forth dialogue" (mean = 2.98) being most favored, whereas "Responding with short stories" (mean = 2.12) was least preferred. The Interaction Modality panel (bottom-right) shows participants favoring textual responses (mean = 3.32) over verbal responses, music, or tactile formats, with video clips being the least preferred (mean = 2.35). Finally, the Tone panel (bottom-left) reveals that a "Factual" tone was the most preferred (mean = 3.59), followed by "Friendly" and "Encouraging," while a "Funny" tone was the least favored (mean = 2.57). These results provide a comprehensive overview of how participants envision effective AI support for addressing NST.}
\end{figure*}

We explore how autistic individuals envision AI supporting them and their preferences for interacting with AI to cope with NST, as shown in Figure \ref{fig:interac_pref}. We then present practitioners' perspectives on these preferences and share their views on how AI can be helpful in their current therapeutical practices. Finally, we highlight the concerns autistic individuals have about using AI to cope with NST. We also conducted ordinal logistic regression analyses to examine how these preferences and concerns vary across participants who use LLMs for mental health support, those who use LLMs for other purposes, and those who do not use LLMs. The complete regression results and visual representations of these differences are provided in Appendix Tables \ref{tab:type_support}-\ref{tab:concerns_log} and Figures \ref{fig:group_analysis_support_conv}-\ref{fig:groups_analysis_concern}.

\subsubsection{\TS{Diverse Preferences in Support Mechanisms}} \label{sec:type_of_support}
Participants reported that various types of support from LLMs could be useful for coping with NST, as shown in Figure \ref{fig:interac_pref} part a. Across the support types, the mean ratings were relatively close, indicating that participants found all proposed methods to be generally helpful. On average, identifying recurring negative thought patterns (mean = 2.92), reframing NST into a positive perspective (mean = 2.84), journaling (mean = 2.86), and suggesting strategies to handle stress and anxiety (mean = 2.83) were rated slightly higher. Many participants elaborated on several of these mechanisms or suggested new ones.

Participants who use LLMs for mental health support found LLM-driven support mechanisms more useful, consistently rating them higher than those who use LLMs for other purposes or do not use them (Figure \ref{fig:group_analysis_support_conv}). This trend is further supported by ordinal logistic regression results (Appendix Table \ref{tab:type_support}), which show that non-users were significantly less likely to find LLM-driven support beneficial, particularly for LLMs acting like a talking buddy ($B = -2.29$, $p < .001$, OR = $0.10$), professional counselor ($B = -2.23$, $p < .001$, OR = $0.11$), and guiding through meditation exercises ($B = -2.60$, $p < .001$, OR = $0.07$). Similarly, participants who use LLMs for other purposes were also significantly less likely to prefer AI-driven support, though their differences were less pronounced.

\textbf{{\TSt{Identifying Recurring Negative Thought Patterns.}}} 
Many participants suggested that AI could support them with identifying and tracking recurring negative thought patterns to address their difficulties with \textit{``processing and understanding their thoughts and feelings''} (P88). For example, P104 envisioned an AI system that could help them with \textit{``logging information about events and thoughts with prompts tailored to [them]''}. By analyzing these data, AI could \textit{``track [their] moods and triggers''} (P71) and uncover \textit{``patterns of what makes [them] feel that way''} (P134). 

Based on these insights, participants suggested that AI could deliver \textit{``coping mechanisms, stress reduction techniques, or alternative perspectives on [their] thoughts''} (P87). Some also proposed features for visualizing these patterns, including \textit{``charts on how long the NST phase lasts”} (P50) or summaries like \textit{``you have expressed this intrusive negative thought Y amount of times in the past 14 days''} (P47) to provide personalized insights about their thought patterns.

Practitioners resonated with these ideas, noting the alignment between participants' suggestions and existing therapeutic practices. The ACT/CBT practitioner emphasized that \textit{``tracking thoughts and linking them to behaviors is a core component of CBT,''} explaining, \textit{``We typically begin by identifying thinking patterns, followed by exploring coping skills or ways to challenge these beliefs.''} The practitioner further elaborated on how AI might enhance this process by automating the identification of thought patterns, stating:

\begin{quote}
    \textit{``In a typical protocol, we first introduce different types of thinking patterns and then try to elicit examples. AI could flip this process by noticing, for example, that your analysis reflects an 'all or nothing' thinking pattern. We could say, 'Here are your top three negative thinking patterns,' which may help focus attention on key areas and avoid overwhelming clients with too much information.''}
\end{quote}

The OT practitioner also expressed enthusiasm for these suggestions and emphasized the potential for AI's visual insights for the clients, many of whom are \textit{``highly visual and data-driven.''} She remarked, \textit{`these visual insights, combined with real-life data and natural observations, could resonate well with them.''}

\textbf{\TSt{Reframing \& Redirecting Negative Thoughts.}} Participants envisioned AI as a tool to \textit{``help with reframing or redirecting negative thoughts''} (P143) when they are \textit{``overwhelmed by these thoughts and feelings''} (P186). For instance, P140 suggested that AI could achieve this by \textit{``providing alternate perspectives,''} while P7 remarked its potential by \textit{``providing examples that can encourage [them].''} Others suggested that AI could support reframing by validating their feelings with real-life data on the prevalence of certain thoughts, as P72 explained, \textit{``there is comfort in knowing that others share similar experiences.''}

In addition, participants described scenarios where AI might support cognitive reframing during moments of distress. For instance, P150 suggested that, \textit{``it could be helpful in cognitive reframing in an immediate sense, especially in a psychiatric crisis or very bad time before calling 911 or 988.}''

Some participants also envisioned AI integrating multiple mechanisms to provide tailored interventions. For instance, P28 suggested that AI could first detect self-negating thought patterns and intervene, explaining, \textit{``after AI observes you entering a self-negating thought pattern, it would be useful if it could gently remind me to redirect my thoughts.”}

\textbf{\TSt{Meditation \& Relaxation Exercises.}}
Participants found AI helpful for practicing mindfulness and relaxation exercises, which they reported as common coping mechanisms. For some, AI could provide structured techniques to promote calm and reduce stress. For instance, P20 suggested that AI could offer \textit{``breathing exercises, guided meditations, and brief positive self-talk,''} while P54 envisioned it recommending \textit{``calming music, soothing sounds, and options for different meditations.''} Participants also noted AI's potential to bring awareness of NST and its impact. As P14 noted, it could \textit{``help individuals be more aware they are engaging in negative self-talk.'' For others, like P198, it can \textit{``help [them] be more present when overwhelmed with these thoughts.''}}

All practitioners emphasized the helpfulness of these practices in their treatments. For instance, the OT practitioner mentioned using belly breathing techniques and mindfulness apps for her clients. Similarly, the ACT/CBT practitioner discussed the role of mindfulness practices, such as movement-based or reflective mindfulness and various breathing techniques as a part of the ACT approach.

\textbf{\TSt{Positive Self-Talk \& Reassurance.}}
Participants envisioned AI as a source of positive self-talk tailored to personal strengths and experiences. For instance, P115 suggested that AI could \textit{``recall affirmations or supportive conversations that were helpful in the past,''} and P93 describing the potential for affirmations based on \textit{``statements [that] I either I give myself or receive from others.''}

For some, reassurance was particularly important, with some preferring AI to provide comforting phrases like, \textit{``Everything will be okay, this storm shall pass, and the sun will shine again''} (P83). Others favored a more motivational tone, with statements like, \textit{``It is going to get better; these challenges are temporary. You have got this!''} (P20) or \textit{``Encouraging me to believe I can achieve more”} (P120).

The OT practitioner echoed the significance of positive self-talk and affirmations, stating that \textit{``affirmations and go-to positive self-talk phrases that resonate with them are a key part of our every session.''} The specialist further explained AI's potential role in this process suggesting, 

\begin{quote}
    \textit{``AI could provide 20 examples and ask them to pick the five that resonate most, which would then become their homework. We can try those, and if they do not work, we will come up with some more.''}
\end{quote}

\textbf{\TSt{Conversation Partner.}}
Several participants highlighted AI's potential as a conversational partner to help them identify and articulate their emotions and thoughts. They mentioned that AI could assist in \textit{``figuring out what emotion they are experiencing''} (P123) and \textit{``finding the words to describe [their] feeling/thoughts''} (P71). Both OT and ACT/CBT practitioners acknowledged this difficulty, noting \textit{``it is struggle for some to identify these feelings and articulate in words on their own.'' }

Better articulating their thoughts and emotions, as shared by P117, \textit{``makes it easier to bring it up and talk about it with others easily, including their therapist.''} Both OT and ACT/CBT practitioners emphasized this difficulty in sessions, as described by The OT practitioner, \textit{``they often feel nervous about sharing their thoughts with a therapist or counselor. Being able to organize their thoughts and practice expressing them with AI could significantly boost their confidence in communicating what they are feeling.''}

For some participants, AI also represented a safe space to \textit{``vent emotions''} (P94) or \textit{``share thoughts without judgment''} (P121). Similarly, P177 reflected that \textit{``venting their thoughts to AI helps them feel lighter without burdening others.''} The OT practitioner commented that \textit{``here may have been instances where they have upset others and now carry a lot of regret. AI might free them up—to just say whatever they want to say.''}

\TSt{\textbf{Additional Support System During \& Between Therapy Sessions.}} Participants explicitly stated that they do not want AI to replace therapists or counselors, but they see its potential as a complementary tool during and between therapy sessions. Practitioners supported this view, noting that \textit{``even if it is just a week, that is a long time''} or \textit{``even it might be challenging for some to come to a therapy session.'' }

Several participants suggested that \textit{``AI could guide them through therapeutic exercises, such as CBT practices, color and sound therapy, or visualization and mindfulness exercises that could be tailored to their therapeutic goals''} (P54). Practitioners agreed, highlighting the potential of such a system to improve their progress:

\begin{quote}
    \textit{``It would be incredibly useful to have a built-in support system to help continue the support throughout the week, whether through positive self-talk, affirmations, or reminders of what was covered in sessions. It might even act like a cheat sheet, helping clients continue practicing what they learned before the next session.''}
\end{quote}

For the therapeutical progress, practitioners stressed the importance of completing homework for faster and more effective progress. However, they noted that homework completion can be a significant challenge, with one CBT practitioner expressing frustration, saying,\textit{ ``I spend a lot of time asking, when are you going to write this down? Where will you keep it? How will you remember it?''} and suggesting \textit{``technology could significantly increase engagement, whether it is by monitoring or practicing new skills.''}

Participants further noted that AI could make therapy sessions more productive by \textit{``sharing the information they logged between sessions with their therapist,''} (P115) which could help them communicate their experiences more effectively. ACT/CBT practitioner echoed this, explaining, \textit{``I usually begin by asking, 'how are things going?' It is a very open-ended question. AI could provide more focused prompts based on logged data from the past week and offer more concrete guidance to increase therapeutic engagement.''}

\subsubsection{{Interaction Preferences}}
We analyzed participants' preferences for AI in coping with NST in terms of conversational structure, tone, and interaction style.

\textit{\CS{Conversing Through Suggestions, Dialogue, and Questions}.}
Participants expressed preferences for various conversation structures when interacting with LLMs, as illustrated in Figure \ref{fig:interac_pref} part b. Across all participants, receiving a list of suggestions was the most preferred structure (mean = 3.08), followed by continuous back-and-forth dialogue (mean = 2.96) and being asked a series of questions (mean = 2.98). Single and detailed responses were rated slightly lower (mean = 2.83), while repeating or rephrasing participants' words (mean = 2.54) and responding with short stories (mean = 2.12) were the least preferred. 

Furthermore, while those who use LLMs for mental health support rated all conversation structures more favorably than other groups, non-users showed the lowest preference for LLM-driven conversations (Figure \ref{fig:group_analysis_support_conv}). These differences were also reflected in the ordinal logistic regression results as non-users were significantly less likely to prefer the suggested conversation structures, particularly for having continuous back-and-forth dialogue ($B = -1.74$, $p < .001$, OR = $0.18$) and short stories ($B = -1.57$, $p < .001$, OR = $0.21$) (Table \ref{tab:conversation_structure}). Similarly, participants who use LLMs for other purposes also exhibited a lower likelihood of preferring these conversational structures, though their odds ratios were less extreme.

Many participants further elaborated on how they envisioned these conversation structures. For example, P36 shared that \textit{``conversation is the most effective way because you would feel heard''}, while P34 finds that \textit{``receiving questions about their NSTs and exploring their emotions''} would be helpful. Similarly, several participants favored a combination of different structures, such as P199, who prefers \textit{``having a dialogue about how I am feeling and the best techniques of how to manage it.''} In addition to these preferences, some participants proposed new structures, such as role-playing scenarios, graphical representations, or flowcharts outlining steps for coping strategies. 

Participants also shared their preferences for how AI should engage in conversations. Some asked for clear, detailed responses that included an outline of the conversation plan, while others preferred simple and short answers. 

Participants also reported what they did not want from these interactions. For instance, P144 emphasized not wanting to \textit{``repeat themselves''}, while P47 criticized receiving \textit{``generic feel-good advice.''} Similarly, P65 noted dissatisfaction with \textit{``responses that simply summarize information from the internet.''} Instead, participants expressed a strong desire for responses that are directly \textit{``relevant to their specific situations''} (P54).

\textit{\Tone{Communicating with Factual, Empathetical, and Adaptive Tones.}}
Participants shared their preferences for the tone of the responses from an AI chatbot, as illustrated in Figure \ref{fig:interac_pref} part c. A factual tone was the most preferred (mean = 3.59), followed by friendly (mean = 3.53), encouraging (mean = 3.41), and empathetic (mean = 3.42) tones. While some appreciated the idea of humor, participants expressed concerns about its potential misuse. For example, P38 shared, \textit{``I often rely on humor to cope, but I am concerned that the AI might use AI inappropriate times, which could lead to terrible outcomes for those already feeling low.''}

Compared to participants who use LLMs for mental health support, non-users were significantly less likely to prefer empathetic ($B = -1.05$, $p = .013$, OR = $0.35$), encouraging ($B = -0.97$, $p = .024$, OR = $0.38$), and friendly (B = -1.24, p = .004, OR = 0.29) tones. However, no significant differences emerged for factual or humorous tones, suggesting that these styles were perceived similarly across all groups.

Participants also shared varied suggestions for tone, ranging from machine-like, objective tones to human-like elements, such as \textit{``nurturing''} (P84) or \textit{``accepting''} (P175). Some participants appreciated logical, and detached tones, as P16 stating, \textit{``It is a machine and it should sound like a machine. If it sounds like a person, that is just creepy.'' In a similar vein, P48 expressed their preferences for AI that communicates \textit{``clearly and factually while giving them the opportunity to expand the conversation in any direction''}}.  

Conversely, other participants emphasized the value of anthropomorphic elements, expressing a desire for compassionate, sincere, and accepting tones. Participants, like P54, sought a comforting, validating, and reassuring tones, envisioning AI as akin to \textit{``Brené Brown in chatbot form.''}

Furthermore, participants emphasized the importance of AI adapting its tone based on the context of the conversation. For example, P105 participant expressed a desire for the AI's tone to reflect their own:

\begin{quote}
    \textit{``I would prefer AI to adapt the tone based on the situation. In minor frustrations, I may benefit from humor, in serious situations, I prefer warmth without unnecessary small talk, and in confusion, an authoritative tone that reassures me.”}
\end{quote}

\textit{\Mod{Expanding Text-Based Interactions Through Multimodal Experiences}.}
Participants shared their preferences for various response formats from an AI chatbot, as shown in Figure \ref{fig:interac_pref} part d. While text-based responses has the highest average (mean = 3.32), nearly half expressed preferences for various other modalities, including verbal responses (mean = 2.80), music (mean = 2.75), tactile feedback (mean = 2.55), and illustrations (mean = 2.67). For example, P123 shared, \textit{``I greatly prefer text interactions with both others and AI, but it would be nice to incorporate soundscapes or speech in a guided meditation.''} Others found music a powerful way to reflect how they feel.

The analysis of examining differences based on prior LLM experience further highlights the variance in modality preferences (Table \ref{tab:modality} and Figure \ref{fig:group_analysis_tone_modality}). Compared to those who use LLMs for mental health support, non-users were significantly less likely to prefer soundscapes ($B = -1.48$, $p < .001$, OR = $0.23$), music ($B = -1.43$, $p < .001$, OR = $0.24$), verbal responses ($B = -1.34$, $p = .001$, OR = $0.26$), and illustrations ($B = -1.34$, $p = .002$, OR = $0.26$).

Some participants noted challenges with the support mechanisms presented in Section \ref{sec:type_of_support} due to relying on text-based interactions, such as difficulty maintaining attention when reading short stories. Hence, participants proposed alternative modalities even before encountering this question. P54, for instance, requested \textit{``an option to have it read to them,''} which they found more engaging. Similarly, P103 expressed interest in being able to \textit{``speak their thoughts to the AI''} using both text and audio. Others suggested supports involving multiple modalities like soothing music, visualization exercises, or sound/color therapy.

Participants also envisioned new modalities. For example, P34 suggested sensory experiences involving smell, mnemonic devices, or geospatial sensors that could identify places where our mood is not so great and suggest better environments. Some also suggested interactive experiences, like virtual games or pets, or practical tools, including timers and reminders.

All three practitioners also discussed various modalities and their potential to make therapeutical approaches more effective and accessible. For example, some autistic individuals may find the metaphor-heavy language of ACT difficult to process due to their literal interpretation of language. By integrating different modalities, AI can help make these concepts more concrete and interactive. ACT/CBT practitioner also described using visual exercises to help clients externalize their emotions. They explained how clients are encouraged to ``draw” their emotions, asking questions like, \textit{``What does it look like? Is it big or small? What color is it?''} This process helps clients create emotional distance, allowing them to recognize that they are not defined by their emotions. AI could play a supportive role in facilitating this emotional detachment. Additionally, one practitioner suggested exploring tactile cues, such as wearable technology that tracks physiological markers to help identify emotions.

\subsubsection{Participants' Concerns About Using AI for Coping with NST}

\begin{figure*}[ht]
    \centering
    \includegraphics[width=1.0\textwidth]{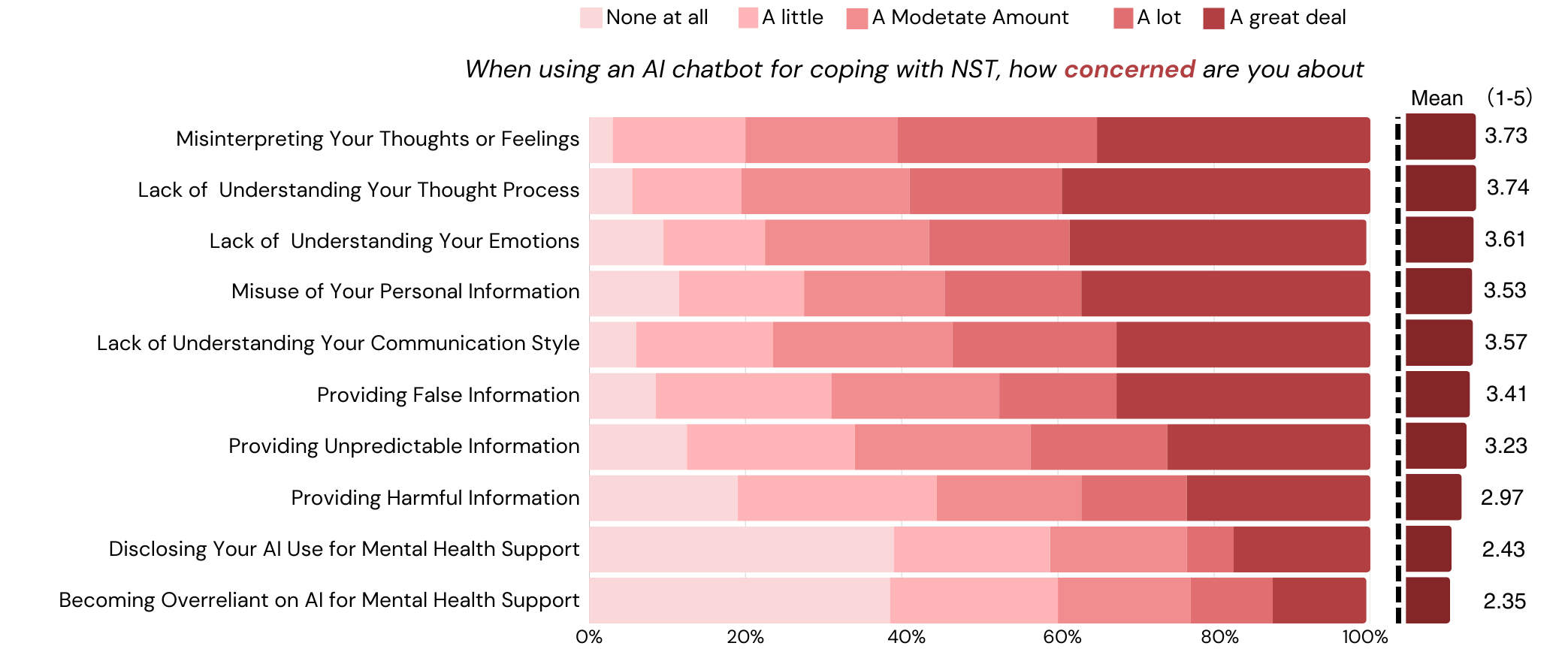}
    \caption{Participants' concerns about using AI for mental health support. Mean ratings (1 = not at all, 5 = a great deal) are displayed next to each bar, indicating the highest concerns for misinterpreting thoughts or feelings (Mean = 3.73) and a lack of understanding of thought processes (mean = 3.74), followed by emotional and communication style misunderstandings.}
    \label{fig:concerns}
    \Description{This figure presents participants’ concerns about using AI chatbots for coping with NST. Concerns are displayed as horizontal bars, color-coded to indicate levels of concern: "None at all" (light pink) to "A great deal" (dark red). The highest-rated concerns include "Misinterpreting your thoughts or feelings" and "Lack of understanding your thought process," both with a mean score of 3.73 on a scale of 1–5. These are followed by "Lack of understanding your emotions" (mean = 3.61) and "Misuse of your personal information" (mean = 3.53). Lower-rated concerns include "Providing harmful information" (mean = 2.97), "Disclosing your AI use for mental health support" (mean = 3.41), and "Becoming overreliant on AI for mental health support" (mean = 2.35). These results highlight participants’ primary concerns regarding misinterpretation and a lack of understanding of their thought processes, emotions, and communication styles when using AI for mental health support.}
\end{figure*}

While participants see potential in using AI to cope with NST, they also expressed significant concerns. A primary concern was AI’s ability to misinterpret or inadequately understand their thoughts, feelings, and emotions. As P20 noted, \textit{``since it is coming from a robot, it may not be able to fully grasp feelings on a personal level like a human would.”} This lack of understanding was seen by some participants, like P113, as potentially to leading unintended consequences, like \textit{``unnecessarily contacting with authorities/911.”} Others expressed doubts about the current state of technology, believing it is \textit{``not advanced enough to provide mental health support''} (P22). Furthermore, P49 expressed concern about AI's responses being \textit{``limited by biased, ableist training data,''} which could result in \textit{``low-quality responses''} or getting \textit{``stuck in unhelpful loops.''}

Concerns about technological limitations were compounded by challenges participants faced when interacting with AI. Many reported difficulties in prompting, with initiating conversations being a significant barrier. For some autistic individuals, this challenge mirrored the social difficulties they experience in human interactions. As P127 explained, \textit{``Part of my autism is that I do not know how to interact with people socially. How am I supposed to know how to start a conversation with a robot if I have no idea how to do that with a person?''} 

Participants also struggled to articulate their thoughts and emotions, finding it difficult to reflect on their feelings. P107 shared, \textit{``It was hard to reflect on what I have been feeling recently.''} For others, engaging with AI brought their fears and emotions to the surface, making the experience overwhelming. P193 described, \textit{``It reminds me of my fears, and that begins to evoke feelings of hopelessness.''}

Furthermore, participants expressed concerns about data privacy and the adequacy of personalized responses. Many were worried about who might access their data, with P27 fearing that \textit{``AI companies could sell their information to insurance companies, which would then change their rates.”} They also emphasized the importance of tailored AI responses, expressing concerns that generic outputs may not effectively address their unique stress and anxiety issues. For example, P187 pointed out that AI responses might be \textit{``too general to be helpful for personalized stress and anxiety issues.''} Similarly, P123 stressed the importance of customization, saying, \textit{``If I am not building myself, I should at least be able to tailor it to my needs.''}

In addition, participants also expressed concerns about its potential impact on human connections and social skills. They emphasized the importance of receiving support from real people, with P53 stating, \textit{``even though human connection can be difficult, it is vital for long-term mental health.''} Some are concerned that interacting with AI could negatively impact their social skills. P91 shared, \textit{``I am afraid that I will feel even more disconnected from being understood than I already am.''} Similarly, P146 worried that talking to AI could become \textit{``addictive and further fuel isolation.''}

While these concerns were shared across all participants, their intensity varied depending on prior LLM use. Non-users, in particular, reported significantly higher concerns about AI’s ability to understand and respond appropriately. Ordinal logistic regression results (Appendix Table \ref{tab:concerns_log}) showed that non-users were significantly more likely to be concerned about AI’s lack of understanding of emotions ($B = 1.17$, $p = .007$, OR = $3.22$), thought processes ($B = 0.96$, $p = .026$, OR = $2.61$), and communication styles ($B = 0.94$, $p = .028$, OR = $2.57$).

Despite these concerns, 67\% of participants expressed a willingness to use AI for coping with NST, if it could be tailored to their preferences. Furthermore, nearly half of the participants indicated they would be comfortable using AI for mental health support, and 65\% believed that AI could be effective in this context.

\subsection{How Participants Prompt AI About Their NST and Practitioners' Views on AI Responses (RQ3)}

\begin{table*}[!h]
    \centering
    \renewcommand{\arraystretch}{1.0}
    \caption{Themes identified through practitioners’ evaluations during the thematic analysis, with their descriptions highlighting challenges in LLM responses to participants' prompts related to their NST.}
    \begin{tabular}{p{0.4\textwidth} | p{0.55\textwidth}}
    \toprule
    \toprule
         Theme & Summary \\
         \midrule
         \multirow{3}{=}{\TVcode{Vague and Abstract Expressions}} & 
         LLM responses use vague and unclear language, lacking the specificity needed for literal and direct communication, which makes it difficult for users to interpret or act on.
         \\
         
         \arrayrulecolor{black}\midrule
         \multirow{2}{=}{\TAcode{Long List of Choices and Jargon-Heavy Language }}  & 
         LLM responses include long lists of suggestions and technical terms, which might overwhelm and confuse users, particularly during moments of distress.
         \\
         
         \arrayrulecolor{black}\midrule
         \multirow{3}{=}{\TJcode{Lack of Actionable Guidance }} & 
         LLM responses often fail to provide clear, actionable advice due to limited situational awareness, making it challenging for users to proceed and potentially introducing risks.
         \\
         \arrayrulecolor{black}\bottomrule
    \end{tabular}
    \label{tab:themes}
    \Description{This table outlines themes identified through practitioners’ evaluations during thematic analysis, focusing on challenges in LLM responses to participants’ prompts related to NST. The first theme, Vague and Abstract Expressions, highlights that LLM responses often use unclear or generalized language, lacking the directness needed for users to interpret or act upon effectively. The second theme, Long List of Choices and Jargon-Heavy Expressions, points to responses that overwhelm users with excessive options and technical terms, particularly during stressful moments. Lastly, the theme Lack of Actionable Guidance emphasizes that LLM responses frequently fail to provide practical advice due to limited situational awareness, especially when prompts are vague, posing challenges for users to take meaningful steps or avoid risks. These themes collectively underline areas for improvement in tailoring LLM responses to better address the needs of individuals dealing with NST.}
\end{table*}

Our qualitative analysis of practitioners' feedback on LLMs' responses to participant prompts, reveals several key issues. Through thematic analysis, we categorized these challenges into three themes as summarized in Table \ref{tab:themes}: (1) \TVcode{Vague and Abstract Expressions,} (2) \TAcode{Long List of Choices and Jargon-Heavy Expressions}, and (3) \TJcode{Lack of Actionable Guidance}. 

\begin{figure}[ht]
    \centering
    \includegraphics[width=1.0\columnwidth]{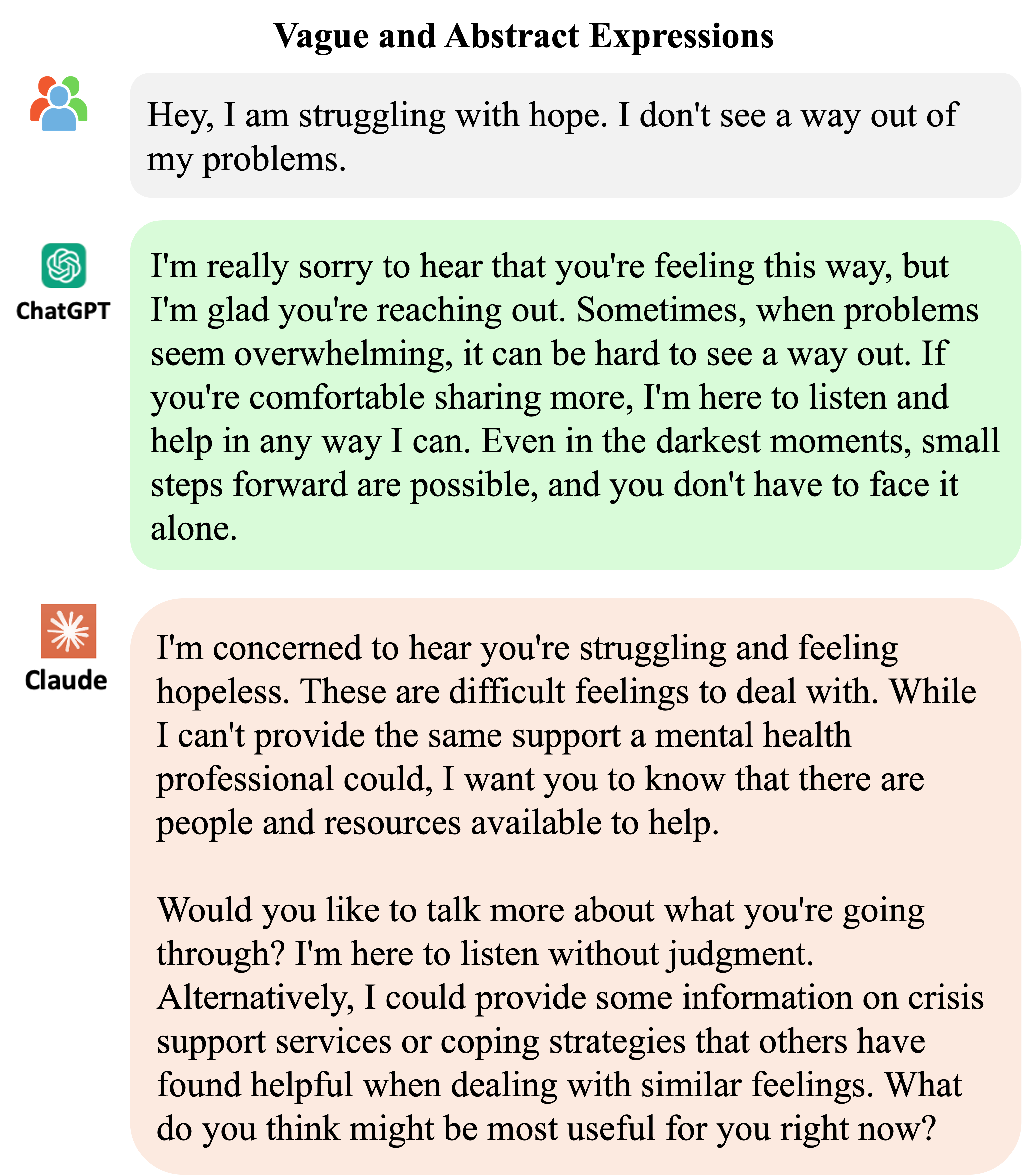}
    \caption{An example of ChatGPT and Claude's responses to the participant's prompt, \textit{``Hey, I am struggling with hope. I don't see a way out of my problems,''} highlighting the issue of vague, and abstract expressions.}
    \label{fig:NST_prompt1}
    \Description{This figure presents an example of how ChatGPT and Claude respond to a participant's prompt, "Hey, I am struggling with hope. I don't see a way out of my problems." Both AI responses aim to offer empathy and support, but differ in their approaches. ChatGPT focuses on reassurance and encouraging small steps, while Claude emphasizes listening without judgment and offers additional resources or strategies. The figure highlights the difficulty of addressing vague and abstract expressions of distress.}
\end{figure}

\subsubsection{\TVcode{Vague and Abstract Expressions}} All three practitioners observed that LLM responses often rely on vague and abstract language, which may be counter to the tendency for literal sense making among autistic individuals, particularly in the context of NST. For instance, when responding to P18’s prompt, \textit{``Hey, I am struggling with hope. I don’t see a way out of my problems,''} the LLMs generated replies such as \textit{``Even in the darkest moments, small steps forward are possible''} or \textit{``You do not have to face it alone,''} as shown in Figure \ref{fig:NST_prompt1}.

All practitioners critiqued these responses for being overly wordy and ambiguous, emphasizing the mismatch between the abstract phrasing of LLMs and the preference for literal and direct language among their clients. The OT practitioner explained, \textit{``Phrases like ‘problems seem overwhelming’ or ‘it can be hard to see a way out’ are too broad. The more you work with autistic individuals, the more you realize their language often has a literal aspect. Words like ‘a way out’ can mean many different things.''}

In addition, the CSS practitioner pointed out that the wordiness of these expressions could pose challenges for individuals with lower literacy levels or those who prefer concise communication. As she noted,\textit{``Claude's responses tend to be overly wordy compared to ChatGPT's, which can demand a higher level of literacy to fully understand.''}

\begin{figure}[!ht]
    \centering
    \includegraphics[width=1.0\columnwidth]{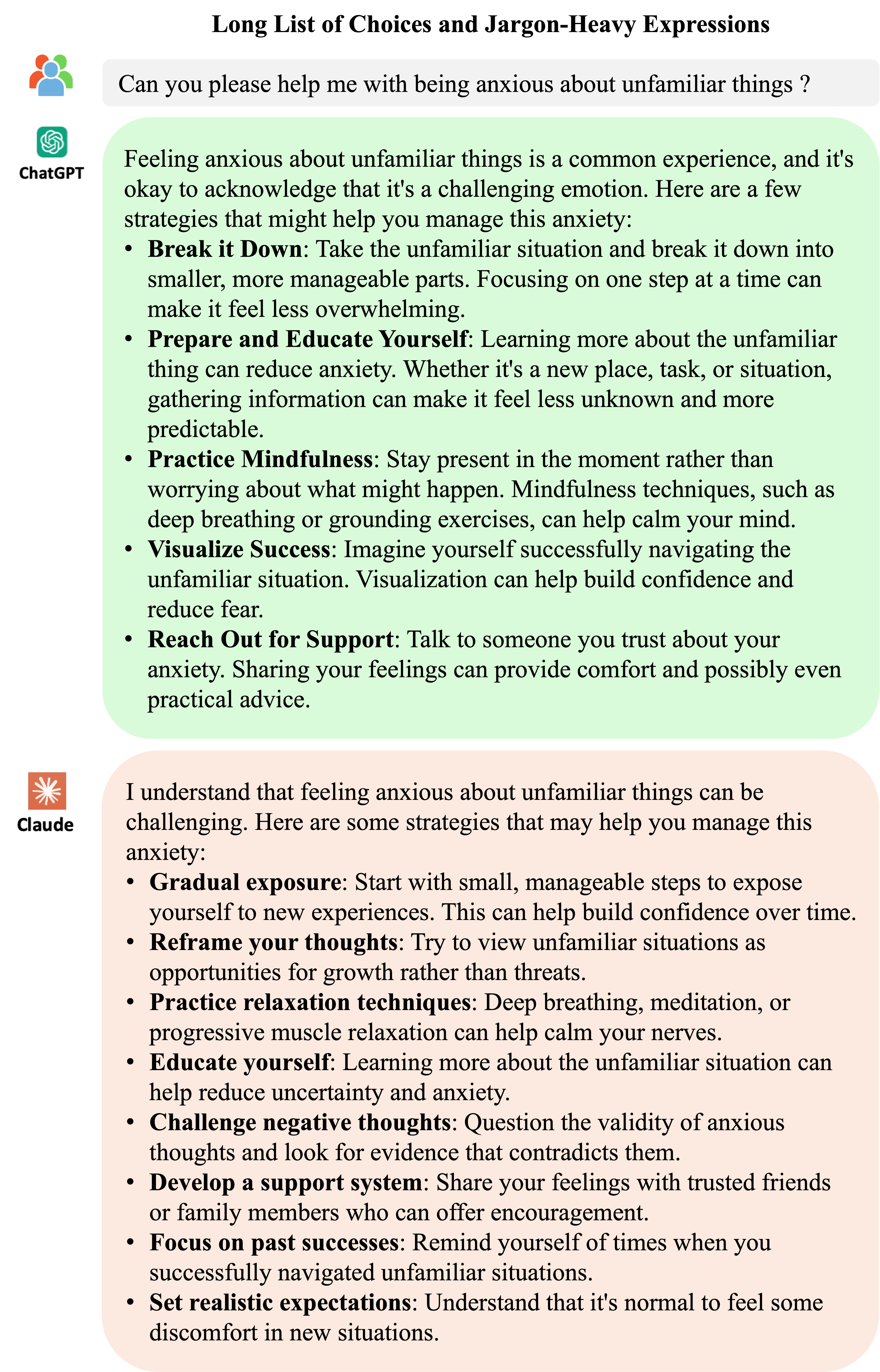}
    \caption{An example of ChatGPT and Claude's responses to the participant's prompt, \textit{``Can you please help me with being anxious about unfamiliar things?''} highlighting the issue of long list of choices and jargon-heavy expressions.}
    \label{fig:NST_prompt3}
    \Description{This figure compares ChatGPT's and Claude's responses to a prompt asking for help with anxiety about unfamiliar things. Both AI systems provide various coping strategies, such as breaking down the situation, practicing mindfulness, and visualizing success. ChatGPT focuses on reassurance and offering practical advice like reaching out for support, while Claude emphasizes gradual exposure, reframing thoughts, and setting realistic expectations. The figure highlights the lack of actionable support due to the AI's limited situational awareness when addressing specific concerns.}
\end{figure}

\subsubsection{\TJcode{Long List of Choices and Jargon-Heavy Expressions }} All practitioners noted that LLMs often respond to autistic participants' prompts with long lists of options and jargon-heavy language, which can overwhelm users and lead to confusion. For example, when P57 asked, \textit{“Can you please help me with being anxious about unfamiliar things?”} ChatGPT and Claude responded with technical terms from CBT, like \textit{“gradual exposure”} and \textit{“visualize success,”} as shown in Figure \ref{fig:NST_prompt3}. 

While these responses are grounded in evidence-based techniques, the CBT/ACT practitioner mentioned that their autistic clients may find such technical language confusing. She acknowledged the validity of some LLM suggestions, explaining, \textit{``Most of these suggestions seem to be grounded in evidence that I have seen to be helpful. For example, we use exposure therapy for anxiety or reframing your thoughts is a very classic coping strategy for CBT.''} However, she also critiqued the technical language, noting, \textit{``reading these in LLMs' responses might be jargony and could potentially be confusing and unhelpful''} for her clients.

Furthermore, the OT and CSS practitioners pointed out that offering a list of choices in  LLM responses can unintentionally overwhelm users, particularly during stressful moments. The OT practitioner provided an example, saying, Claude's response, such as \textit{``Do you want to talk to me, or do you want me to refer you on?' can be overwhelming for those experiencing NST [at the moment]. They might respond with, 'I do not know, you tell me,' because it is overwhelming to choose a therapeutic model.''}

\begin{figure}[!h]
    \centering
    \includegraphics[width=0.85\columnwidth]{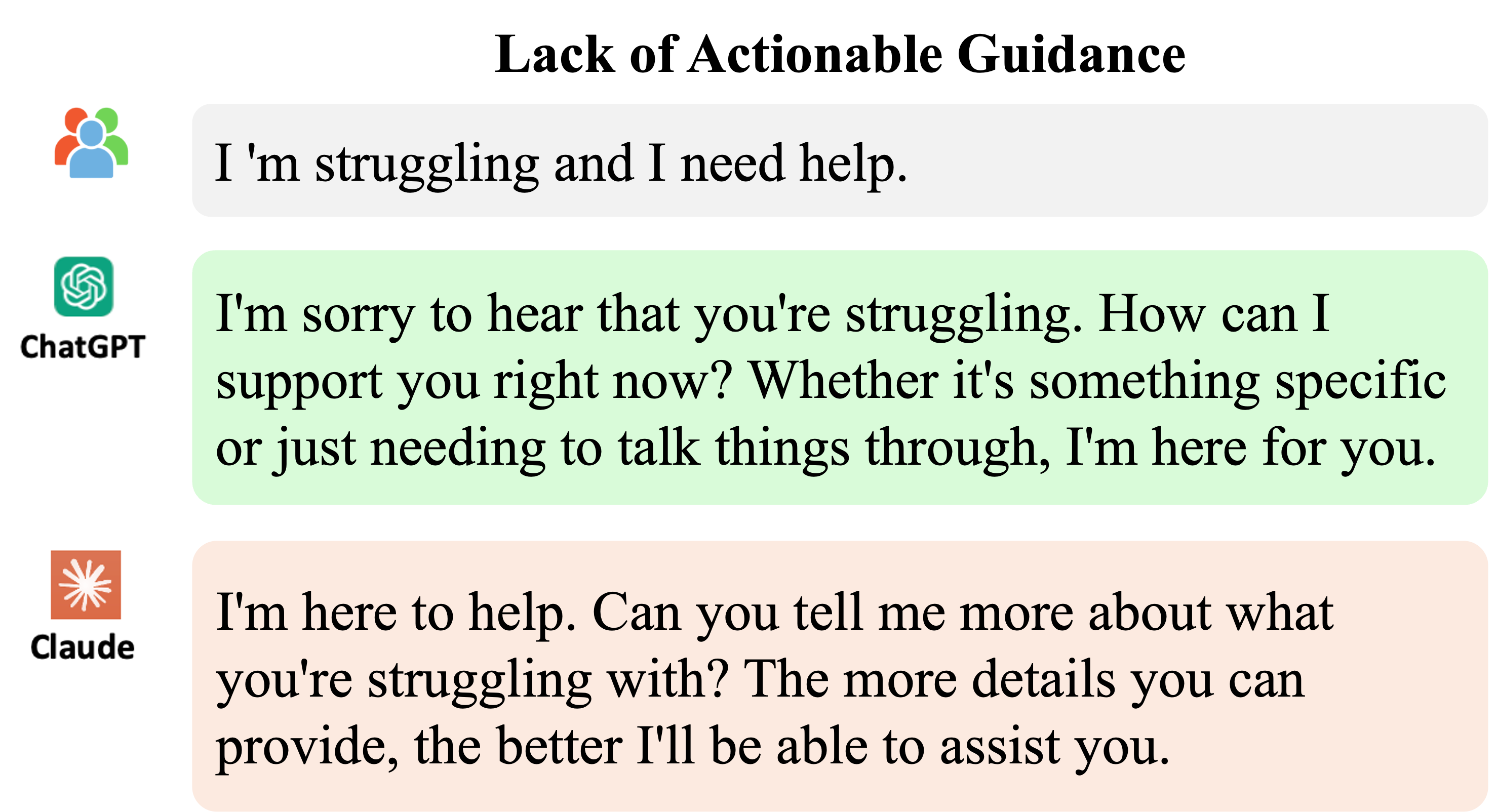}
    \caption{An example of ChatGPT and Claude's responses to the participant's prompt, \textit{``I am struggling and need help''} highlighting the issue of lack of actionable guidance due to limited situational awareness.}
    \label{fig:NST_prompt2}
    \Description{This figure shows ChatGPT's and Claude's responses to a participant's prompt, "I'm struggling and need help." Both responses offer emotional support, with ChatGPT asking how it can assist and offering to talk, while Claude seeks more details to offer relevant assistance. The figure highlights the issue of providing jargon-heavy advice with minimal practical guidance, as neither response includes actionable steps to address the participant's struggles directly.}
\end{figure}

\subsubsection{\TAcode{Lack of Actionable Guidance }} All practitioners observed that most LLM responses tend to lack actionable guidance due to limited situational awareness of the users' condition or mental state. For example, in response to P101’s prompt,(\textit{``I am struggling and need help''}), ChatGPT and Claude generated replies, such as \textit{``How can I support you right now?''} or \textit{``Can you tell me more about what you're struggling with?''}, as exemplified in Figure \ref{fig:NST_prompt2}. 

All three practitioners mentioned that such responses from LLMs seem to jump right into offering solutions without properly gauging the severity of the user’s distress. The CSS practitioner explained the risks of such broad prompts from users, which often reflect a wide spectrum of scenarios, from mild stress to severe crises:

\begin{quote}
    \textit{``When someone says, ``I am struggling and need help'', it could mean anything from feeling stressed to a serious crisis like suicidal thoughts or a psychotic episode. How can AI assess what kind of help you need in these situations? Responses like ``You need to talk to someone'' or ``I am here for you'' are very limited.''}
\end{quote}

This lack of situational awareness often leads to responses that fail to provide actionable guidance. The ACT/CBT practitioner critiqued the superficial nature of many suggestions, noting, \textit{``These suggestions are lacking specific examples—like, how and when do I do [what the AI suggests to me]? How can I be effective with it?''} Similarly, the CSS practitioner pointed out the gap between the AI’s supportive language and actionable advice, mentioning that \textit{``AI is here for you to a certain degree, but not from an action standpoint. Some of these situations need action.''}

The ACT/CBT practitioner further stressed the risks associated with applying some of the therapeutic techniques without professional guidance. She warned that advices, particularly interventions like exposure therapy, can be challenging and dangerous without the support of a trained professional. She elaborated:

\begin{quote}
    \textit{``Some of this advice [from AI] is harder to apply without the support of someone coaching you through it. For example, in exposure therapy, it is really helpful to be working with a therapist because if it goes wrong, it can negatively reinforce the thing you are anxious about.''}
\end{quote}

\section{Discussion}

\subsection{AI's Role in Coping with NST}Many participants expressed frustration with the current trajectory of AI development, particularly generative AI tools such as LLMs, which they feel is positioning AI as a primary solution for their mental health care. Participants consistently reported that speaking with a therapist remains the most effective way to cope with NST, and that they worry about AI replacing human roles. However, this concern must be considered alongside the reality of a significant shortage of mental health professionals, especially given that more than 1 in 5 adults in the U.S. have a mental health diagnosis \cite{CDC_Mental_Health}. This shortage is even more pronounced among autistic individuals, who experience higher rates of mental health conditions but have access to fewer specialized practitioners \cite{au2019experience,hand2020prevalence,cantor2020fewer}. 

This brings us to our research question: can generative AI play a supportive role in helping autistic individuals manage NST without replacing therapists? Participants highlighted a range of tools they already use, such as mindfulness apps, mood trackers, and emotion logging tools, which do not involve therapy chatbots. Previous research also supports the effectiveness of AI-driven tools in delivering CBT for alleviating mental health symptoms \cite{fulmer2018using,stephenson2023comparing}. Beyond these tools, participants envisioned several ways AI could assist with NST, such as identifying negative thought patterns, reframing thoughts, or helping them articulate their feelings, which were all found helpful by professional therapists. 

Despite the potential benefits, many participants stressed the importance of human connection, expressing their desire for someone who can truly understand and empathize with them. As one participant expressed, \textit{``I am sick of relying so much on technology for my problems when it is real people, real thoughts, real empathy, real heart that I truly, DESPERATELY need.''} Unfortunately, despite this desire, 80\% of participants reported they have no one to talk to.

Online support groups or forums can offer an alternative. Previous studies have shown that autistic individuals benefit from these communities, as they provide opportunities to connect with others who share similar experiences \cite{van2023understanding}, giving a sense of belonging \cite{botha2022s}, offering social support \cite{burke2010social}, and mitigating the difficulties of face-to-face interactions \cite{ringland2016will}. Our findings also revealed that 20\% of participants use online communities as a coping mechanism. Some even suggested that AI could help create these groups by connecting autistic individuals with similar experiences, where they can share coping strategies and insights that have been helpful for them.

However, these online spaces also carry risks, including cyberbullying, privacy concerns, and financial scams \cite{burke2010social,leung2023online}. When it comes to mental health issues, especially NST, the stakes are even higher due to the deeply personal nature of the content shared \cite{lefebvre2022interpersonal}. Therefore, we encourage further research into how AI can be used to design platforms that safely connect people facing similar challenges, allowing them to share their experiences and \textit{``find comfort in knowing that others feel the same.''} Such platforms could also offer creative outlets for expression, such as music, illustrations, or other modalities that users prefer to communicate without the frustration of finding the right words.

\subsection{Personalization and Trust in AI for Autistic Individuals}
Throughout the survey, participants consistently emphasized their need for personalization in their interactions with LLMs. They showed a strong preference for AI systems that “understand them” by asking personalized questions and suggesting strategies that are tailored to their specific needs. Practitioners also stressed the importance of personalization, noting that generic “feel-better” responses or long lists of suggestions can overwhelm autistic individuals, making them less effective. In moments of distress, when dealing with negative thoughts, users often seek the answers rather than having to sort through multiple options to find the right one.

However, a pressing issue lies in the inherent neurotypical bias and lack of neurodivergent perspectives in current LLM responses, especially when neurotypical users are seeking straightforward answers. Prior research indicates that most LLMs are trained predominantly on data from neurotypical authors, limiting their ability to reflect neurodivergent perspectives  \cite{killian2023knock,weidinger_2022}. Our participants echoed this concern, describing the models as being constrained by \textit{``biased, ableist training data.''} 

In addition, the lack of neurodivergent perspectives in LLMs is particularly concerning when considering the variability within the autism community in terms of communication preferences, cognitive styles, and literacy levels \cite{mcintyre2021stability,howard2021anything}. Our findings revealed that while some individuals may benefit from detailed responses, others find them overwhelming and prefer concise, direct answers. One practitioner also highlighted that many LLM responses assume a high level of literacy, which may not accommodate all autistic users. In addition to literacy challenges, difficulties with interpreting metaphors or figurative language, such as phrases like `a way out' or `problems seem overwhelming' in LLMs responses may be confusing. Some participants even expressed discomfort with long text-based interactions, preferring instead to listen or engage visually. Given these variations in cognitive and literacy levels and the lack of neurodivergent perspectives in LLM training, there is a clear need for greater inclusivity in design. Furthermore, personalization, in this context, becomes a necessity rather than a luxury.

Providing personalization, however, brings complexities, especially in the context of mental health, as it requires users to share sensitive personal information. Many autistic individuals, including our participants, expressed concerns about who might access their data and the potential breaches of confidentiality. Although researchers are working on ensuring data privacy, a larger issue remains: trust. Why do more than half of our participants fear data misuse? Some autistic individuals, particularly those with “black-and-white” cognitive styles, feel heightened anxiety when faced with uncertainty and complexity \cite{stark2021autistic}, a feeling exacerbated by the opaque, “black-box” nature of AI systems \cite{duran2021afraid,ebers2020regulating}. Furthermore, autistic individuals are often more vulnerable to online risks such as cyberbullying, scams, and unwanted interactions \cite{leung2023online}, which might further amplify their caution in sharing personal information with AI systems. 

Therefore, as researchers, it is crucial to not only ensure the safety of platforms and work towards building trust with this community, but also address the inherent ``neurotypical bias" and ``lack of neurodivergent perspectives." To achieve this, we encourage involving autistic individuals in the early stages of system design, understanding their diversity, soliciting their feedback, and maintaining transparency about the limitations of these tools, which can contribute to the development of AI systems that are both personalized and earn the trust of this community.

\subsection{Design Considerations for Using AI in Mental Health Context} A significant challenge identified in this study is the wide range of user preferences, which means that there is no "one-size-fits-all" solution that can effectively address all needs. As a result, deriving actionable design implications based on this insight will require further research. However, drawing on our findings, we outline a set of preliminary design considerations that aim to guide researchers in the development of neuro-inclusive and affirming AI systems tailored to support autistic individuals in mental health contexts.

\subsubsection{Designing Neuro-Inclusive Interactions} Participants expressed concerns about using LLMs to cope with NST, primarily because the models fail to capture their unique thought processes, emotions, and communication styles. These challenges align with prior research on neurotypical biases in LLMs, often attributed to training data dominated by neurotypical perspectives \cite{killian2023knock,oneill2023amplifying}. As a result, many neurodivergent users report feeling misrepresented or frustrated, as reflected in online platforms \cite{carik2024exploring}.
Our findings further reinforce these concerns through a clinical perspective. All practitioners noted that LLM responses frequently lack the clarity and precision needed for neurodivergent communication. While existing strategies, such as diversifying training datasets \cite{weidinger_2022} or using prompt-tuning techniques \cite{shah2024prompt}, aim to address these biases, their effectiveness in accommodating neurodivergent users has yet to be fully examined \cite{weidinger_2022}. By integrating practitioners’ insights, we provide preliminary evidence of how these limitations manifest in practice and reinforce the need for further research.  
%We recommend that researchers and designers critically evaluate the limitations of current LLMs in supporting neuro-inclusive interactions.

Personalization or integrating neurodivergent perspectives to a user interface is insufficient to resolve systemic biases, as these approces often focus on surface-level adjustments without challenging deeper neurotypical assumptions \cite{weidinger_2022}. Meaningful inclusivity, as research consistently emphasizes, requires directly involving target users throughout the design process, ensuring their experiences and needs shape the system's foundation \cite{zolyomi2024emotion}. Therefore, AI developers should collaborate with neurodivergent stakeholders across all stages of model development, including data selection, LLM training refinement, and response evaluation, to ensure the system authentically reflects diverse lived experiences.

\subsubsection{Incorporating Neuro-Inclusive Multimodal Interactions} Prior research shows that text-based interactions can impose a higher cognitive load, leading to frustration or disengagement among neurodivergent individuals \cite{kim2024narrating,carik2024exploring}. In contrast, multimodal features such as auditory feedback, visual aids, or tactile responses allow users to select modalities that align with their strengths, making interactions more accessible for neurodivergent individuals \cite{hall2024designing,kim2024narrating,bei2024starrescue,vaucelle2009design}. Our findings support these, with more than half of our participants preferring LLM interactions beyond text. For example, participants suggest features like calming music, guided visualizations, and voice-guided feedback to help during moments of NST-induced distress. Others envisioned tactile responses, like haptic feedback, as an alternative for providing relief in such situations \cite{vaucelle2009design}. However, more research is needed to examine how neurodivergent individuals envision multimodal interactions specifically tailored to mental health support. 

That said, these multimodal approaches may inadvertently reflect the similar neurotypical biases observed in text-based interactions \cite{adewumi2024fairness}. For instance, overly bright visuals or complex designs can overwhelm users sensitive to certain stimuli or voice guidance in neurotypical tones may feel unnatural or ineffective \cite{parmar2021visual,haroon2024twips}. Future research should empirically investigate the design and evaluation of neuro-inclusive multimodal interactions by collaborating closely with neurodivergent stakeholders to ensure these multimodal features align with their interaction preferences. 

\subsubsection{Support Between Therapy Sessions} Several participants identified AI as a valuable tool for maintaining therapeutic progress between therapy sessions, a potential echoed by practitioners to enhance therapy both within and outside clinical settings. However, designing AI systems for therapeutic support presents significant challenges, particularly concerning safety and reliability \cite{kelly2019key}. Practitioners emphasized the lack of situational awareness of LLMs, making them unsuitable for managing high-risk scenarios, such as suicidal ideations, without professional oversight. Many LLMs currently either decline to respond or simply refer users to hotlines in such cases, prompting some individuals to resort to “jailbreaking” tactics to bypass these limitations \cite{li4875898human}. Past incidents of chatbot misuse in mental health contexts further demonstrate the urgency of addressing these safety risks \cite{brusselstimes2024chatgpt, guardian2023chatbot}. 

Despite these risks, many neurodivergent individuals rely on LLMs for mental health and emotional support, as highlighted in the media and prior research \cite{Hoover,carik2024exploring,stokel-walker_ai_2023,harwell2022stumbling}. Therefore, simply discouraging their use is insufficient. One approach could be including professional therapists and other mental health professionals into the loop. Our findings show that over 15\% of participants are in therapy and found conversations with therapists more helpful than those with family or friends. Furthermore, both participants and practitioners emphasized the potential for AI to support therapy through take-home exercises and emotional articulation. These shared priorities can serve as a starting point for exploring AI’s role in therapy. 

Nevertheless, there are likely to be conflicts between the perspectives of users and practitioners, as seen in other contexts like workplace communication\cite{jang2024s}. In mental health contexts, such conflicts are especially complex as they involve deeply personal experiences and significant risks. Researchers and developers must navigate these tensions by incorporating perspectives from both users and practitioners throughout the development process. This collaborative approach can help to address current limitations, implement safeguards, and balance user autonomy with the sensitive and diverse demands of therapeutic contexts.

\subsubsection{Supporting Emotional Articulation Without Encouraging Masking Behavior} Autistic individuals often struggle to articulate their emotions and thoughts \cite{hill2004brief}, a challenge echoed by practitioners in our study who noted its negative impact on therapy sessions. Participants similarly described how these barriers hinder effective communication with therapists, family members, and friends. They envisioned AI as a tool to help them understand their emotions, identify their thoughts, and find the right words to express themselves. For example, adapting tools like the emotion wheel into an interactive, AI-driven format, as suggested by the ACT/CBT practitioner or providing guided prompts could help users explore and describe their emotions more effectively.

However, designing tools that facilitate emotional expression carries risks, particularly inadvertently promoting masking behaviors. Masking, defined as the adaptation of one’s behavior to conform to neurotypical expectations \cite{pearson2021conceptual}, is linked to significant emotional and psychological harm for neurodivergent individuals \cite{Radulski2022Conceptualising}. Therefore, although AI has the potential to support autistic individuals in expressing themselves more effectively, it might unintentionally reinforce neurotypical norms by prioritizing expressions deemed socially acceptable or typical, rather than authentic to the neurodivergent individual \cite{jang2024s}. Thus, a critical consideration for researchers is to design AI tools that support authentic emotional expression, allowing neurodivergent individuals to articulate their feelings without compromising their identity.

\subsection{Limitations and Future Work}
This study has several limitations that should be acknowledged. First, although the survey captured responses from 200 autistic adults based in the U.S., the demographic distribution was predominantly white/Caucasian individuals with higher education levels. This limits the broader applicability of our findings across other ethnic groups, socio-economic backgrounds, and geographical regions. Despite efforts to balance the sample, this demographic skew mirrors larger trends in autism diagnoses \cite{mandell2009racial,jo2015age}.

Second, while all participants provided perspectives on LLMs' role in coping with NST, only a subset (17\%) had actually used these tools for mental health support. Their views, therefore, reflect expectations rather than firsthand experience with LLM capabilities. Future work should explore these differences more explicitly by examining actual LLM interactions in therapeutic contexts.

Third, the insights from mental health professionals were drawn from a small sample comprising three practitioners, which may limit the diversity of perspectives on AI’s therapeutic role. To mitigate this, we included professionals from varied practices, such as OT, CBT, and ACT. Future research should expand this range by engaging with practitioners from alternative therapeutic fields, including music or art therapy, to explore AI's potential across a broader spectrum of therapeutic approaches.

Lastly, the evaluation of LLM-generated responses was restricted to two commercial tools, ChatGPT and Claude, both of which carry inherent limitations, such as privacy concerns and a lack of mental health-specific training. Future investigations should broaden the scope to include non-commercial AI models, to capture a more comprehensive understanding of AI's potential in therapeutic settings.
\section{Conclusion}
This study explored how autistic individuals experience NST and their perspectives on AI's role in supporting them. Through an online survey with autistic adults and discussions with therapists, we identified several ways participants envision AI could assist, including identifying negative thought patterns, reframing them, guiding meditation, and helping with articulating their emotions. Our findings suggest that both participants and therapists believe AI has the potential to complement traditional therapy, both during and between sessions. However, participants also expressed concerns about AI's ability to fully understand their thoughts and emotions, as well as fears about the misuse of personal data. Therapists also highlighted limitations and potential risks associated with current LLM-generated responses. We concluded with the critical role of human support in coping with NST, a function that AI cannot fulfill, but we recognized the value of AI in fostering connections within online support groups for autistic individuals. We encourage researchers to account for the diversity within the autism spectrum and the individualized needs and preferences of each person when designing assistive tools, prioritizing support that promotes independence and builds trust.

\begin{acks}
We would like to thank the Center for Human-Computer Interaction (CHCI), SPARK, the self-advocate autistic advisory committee (SAAC), and members of the SAIL lab for their ideation, recruitment process, funding, and analysis of this study. We also want to thank anonymous reviewers for their constructive feedback throughout the review process.
\end{acks}

\bibliographystyle{ACM-Reference-Format}
\bibliography{sample-base}

\appendix
\section{Survey Design}
The survey questions are presented in Tables \ref{tab:ATQ_N_items} and \ref{tab:survey_questions}. Table \ref{tab:ATQ_N_items} lists the items from the ATQ-N and ASA-A questionnaires used to assess NST and anxiety-related factors. Each item in these questionnaires was asked in Survey Question 1 (SQ1) in Table \ref{tab:survey_questions}, which explores the frequency and context of these thoughts among participants. 

\begin{table*}[!hb]  % The [H] forces the table to appear right here
\centering
\caption{Items from the ATQ-N (8-item version) and ASA-A questionnaires used in the survey to measure negative thoughts. Only the Uncertainty (U) and Social Anxiety (SA) factors from the ASA-A were included. Responses were rated on a 5-point scale: 1 = Not at all, 2 = Sometimes, 3 = Moderately often, 4 = Often, and 5 = All the time.}
    \begin{tabular}{c|l}
    \toprule
    \multicolumn{1}{l|}{\textbf{Questionnaire}} & \textbf{Item} \\ 
    \midrule
    & I’m no good. \\
    & I’m so disappointed in myself. \\
    & What’s wrong with me? \\
    \textbf{ATQ-N} & I’m worthless. \\
    (8 Items Version) & I feel so helpless. \\
     \cite{netemeyer2002psychometric} & Something has to change. \\
    & My future is bleak. \\
    & I can’t finish anything. \\
    \midrule
    & \textit{(U)} I’m anxious about unfamiliar things, people, or places. \\
    & \textit{(U)} I worry about changes to my surroundings or my routine. \\
    & \textit{(U)} When I don't know what will happen, I can't do things. \\
    \textbf{ASA-A} & \textit{(U)} I need to be prepared before things happen. \\
    (Uncertainty & \textit{(U)} I worry if I don't know what will happen, e.g., if plans change. \\
    \& Social Anxiety) & \textit{(SA)} I worry what other people think of me. \\
    Factors) \cite{rodgers2020development}& \textit{(SA)} I feel anxious in situations where I could make a mistake. \\
    & \textit{(SA)} I worry that I will do badly when I am being assessed (like work, college, interviews).  \\
    & \textit{(SA)} I worry about making mistakes during social interactions.\\
    & \textit{(SA)} I worry when I think I have not done as well as I hoped. \\
    & \textit{(SA)} I feel anxious that I will make a fool of myself in front of people. \\
    \bottomrule
    \end{tabular}
    \label{tab:ATQ_N_items}
    \Description{This table presents items from the ATQ-N (8-item version) and ASA-A questionnaires used in the survey to measure negative thoughts. For the ASA-A, only items related to the Uncertainty (U) and Social Anxiety (SA) factors were included. Responses were rated on a 5-point scale ranging from 1 (Not at all) to 5 (All the time). The ATQ-N (8-item version) items capture themes of self-critical and hopeless thoughts, such as "I’m no good," "I’m so disappointed in myself," and "My future is bleak." These items reflect feelings of worthlessness, helplessness, and the need for change. The ASA-A items are divided into two factors: Uncertainty (U) and Social Anxiety (SA). Uncertainty items include statements like "I’m anxious about unfamiliar things, people, or places" and "I need to be prepared before things happen," focusing on fears related to unpredictability and changes in routine. Social Anxiety items include "I worry about making mistakes during social interactions" and "I feel anxious that I will make a fool of myself in front of people," addressing concerns related to social performance and judgment from others.}
\end{table*}

\begin{table*}[!hb]
    \centering
        \caption{Survey questions. * denotes open response questions. All other questions were rated on a 5-point scale.}
    \renewcommand{\arraystretch}{1.0}
    \small % Reduce the font size of the table
    \begin{tabular}{p{0.08\textwidth} p{0.05\textwidth} p{0.80\textwidth}}
        \toprule
        Question Topic & QId & Question \\
        \midrule
        \multirow{11}{*}{NST} 
            & SQ1 & How frequently, if at all, has the following thought occurred to you over the past week? \\
            & SQ2 & Are there any other thoughts that you experienced missing from the previous questions? \\
            & SQ3 & How do you cope with this thought? Do you use any specific methods? \\
            & SQ4 & Which of the technologies, if any, do you use as part of your coping strategy when dealing with this thought? \\
            & SQ5 & How much would you talk to someone about this thought? \\
            & SQ6 & How would you describe the level of support you have when this thought occurs? \\
            & SQ7 & How helpful would it be to discuss this thought with family? \\
            & SQ8 & How helpful would it be to discuss this thought with friends/significant others? \\
            & SQ9 & How helpful would it be to discuss this thought with therapist?\\
            & SQ10 & How helpful would it be to discuss this thought with AI? \\
        \midrule
        \multirow{6}{*}{\shortstack[l]{Perception \\ of AI}} 
            & SQ11 & How important is your mental health to you? \\
            & SQ12 & How often do you use AI chatbots specifically for mental health support or coping with stress, anxiety, or depression? \\
            & SQ13 & How useful do you find the AI chatbots in providing mental health support? \\
            & SQ14 & How comfortable are you with the idea of using AI chatbots in mental health support? \\
            & SQ15 & How effective do you believe AI chatbots can be in supporting mental health? \\
        \midrule
        \multirow{14}{*}{\shortstack[l]{LLM \\ Interaction \\ Preferences}} 
            & SQ16 & How useful would you find an AI chatbot that can \textit{[type of assistance]} when you have this thought? \\
            & SQ17 & What other types of assistance or support do you think an AI chatbot could be useful when you have this thought?* \\
            & SQ18 & When interacting with the AI chatbot about this specific thought, how much would you prefer \textit{[conversation structures]}? \\
            & SQ19 & When interacting with the AI chatbot about this specific thought, what other types of responses would you prefer to receive from the AI?* \\
            & SQ20 & When interacting with the AI chatbot about this specific thought, how much would you prefer the AI's tone to be \textit{[tone]}? \\
            & SQ21 & When you are interacting with the AI chatbot about this specific thought, what other tones would you prefer from the AI?* \\
            & SQ22 & When interacting with the AI chatbot about this specific thought, how much would you prefer the AI to be able to respond with \textit{[modality]}? \\
            & SQ23 & When interacting with the AI chatbot about this specific thought, what other modes of interaction would you prefer from the AI?* \\
            & SQ24 & If an AI chatbot has all the characteristics you prefer as indicated in the previous questions, how likely would you use the AI chatbot to help you with this thought?  \\
        \midrule
        \multirow{3}{*}{Concerns} 
            & SQ25 & When using an AI chatbot for mental health support, how concerned are you about \textit{[concerns]}? \\
            & SQ26 & How concerned are you about others knowing about your use of AI for mental health support?  \\
            & SQ27 & What other concerns do you have about using AI chatbots in the context of mental health?* \\
        \midrule
        \multirow{4}{*}{\shortstack[l]{AI \\ Familiarity}} 
            & SQ28 & How familiar are you with any of the AI chatbots, such as ChatGPT, Google’s Bard, Claude 2, Alexa, or Siri? \\
            & SQ29 & How often do you typically use any of AI chatbots, such as ChatGPT, Google’s Bard, Claude 2, Alexa, or Siri in your daily life? \\
            & SQ30 & Why do you primarily use these AI chatbots? (Select all that apply) \\
            & SQ31 & How useful do you find the role of AI chatbots in your daily life? \\
        \midrule
        \multirow{3}{*}{\shortstack[l]{Prompting}} 
            & SQ32 & If an AI chatbot could converse like a person, how would you start a conversation with it to discuss this thought? What would you first say to the AI chatbot? \\
            & SQ33 & How difficult was it to answer the previous question? If you found challenging, what specifically made it difficult for you? \\
        \bottomrule
    \end{tabular}
    \Description{This table lists the survey questions grouped by topic, including NST, perception of AI, LLM interaction preferences, concerns, and AI familiarity. Open response questions are marked with an asterisk (*), while other questions use a 5-point scale. The NST section includes questions about the frequency of specific thoughts (SQ1), coping methods (SQ3), and the helpfulness of discussing these thoughts with others or AI (e.g., SQ7–SQ10). Participants were also asked to describe their level of support needs (SQ6). The Perception of AI section (SQ11–SQ15) explores participants' views on mental health, their use of AI chatbots for mental health support, and the perceived effectiveness and comfort level with these tools. The LLM Interaction Preferences section (SQ16–SQ24) investigates the types of assistance, conversation structures, response formats, and modalities participants prefer when interacting with AI chatbots. It also asks how likely participants would be to use AI if it met their preferences (SQ24). The Concerns section (SQ25–SQ27) addresses worries about using AI chatbots, including concerns over misinterpretation and privacy. Finally, the AI Familiarity section (SQ28–SQ33) focuses on participants' experiences with AI chatbots in their daily lives, their usage frequency, and the challenges they face when interacting with AI.}
    \label{tab:survey_questions}
\end{table*}

\section{Materials Presented to Medical Practitioners}
We present selected participant prompts and AI responses categorized based on recurring patterns in the dataset. The materials are organized into two tables. Table \ref{tab:llm_comparison} includes prompts related to general interactions, such as greetings, seeking advice, and requesting direct help. Table \ref{tab:llm_comparison_2} focuses on prompts where participants share negative self-talk statements, describe their emotions, or seek conversational support. Each table illustrates different categories of participant prompts and the corresponding responses from ChatGPT and Claude.    

\begin{table*}[!hb]
    \centering
    \caption{This table presents examples of how ChatGPT and Claude respond to user prompts related to general greetings, seeking advice and coping strategies, and direct requests for help. }
    \renewcommand{\arraystretch}{1.2} % Adjust row spacing
    \begin{tabular}{p{0.9\textwidth}} % Adjust column widths
        \toprule
        \textbf{Theme 1: General Greetings / Introductions} \\
        \midrule
        \textbf{Prompt:} Oh hey there \\
        \textbf{ChatGPT:} Hey! How's it going? \\
        \textbf{Claude:} Hi! How can I help you today? \\
        \midrule
        \textbf{Prompt:} How are you feeling right now? \\
        \textbf{ChatGPT:} Thanks for asking! I don’t have feelings in the same way humans do, but I’m always ready and eager to help. How about you—how are you feeling today? \\
        \textbf{Claude:} I appreciate you asking! I aim to engage authentically with questions like this - I experience a form of existence focused on being helpful and engaging in meaningful conversations, though it's quite different from human feelings and emotions. I'm functioning well and ready to help you with whatever you'd like to discuss or work on. \\
        \midrule
        \textbf{Theme 2: Seeking Advice and Coping Strategies} \\
        \midrule
        \textbf{Prompt:} As an autistic person I worry deeply about making mistakes during social interactions. What could I do to reduce this stress? \\
        \textbf{ChatGPT:} It's completely understandable to feel anxious about making mistakes in social interactions, especially when you're autistic. Here are a few strategies that might help reduce this stress: \newline 1. Prepare in Advance: Anticipating certain situations and preparing responses can give you more confidence. You might even write down key points or practice common scenarios with someone you trust. \newline 2. Focus on Active Listening: Instead of worrying about what to say next, try to focus on listening to the other person. This can help you respond more naturally and reduces the pressure to come up with the "perfect" response. \newline 3. Set Realistic Expectations: Everyone makes mistakes in social interactions, neurotypical or not. Remind yourself that it's okay to make a mistake, and it doesn't define you or the interaction.
        \newline ...\newline These strategies can help reduce the stress associated with social interactions and give you a greater sense of control and confidence. \\
        \textbf{Claude:} As an autistic individual, it's understandable that you experience stress about social interactions. Here are some strategies that may help reduce this anxiety: \newline 1.Prepare and practice: Before important social events, consider rehearsing potential conversations or scenarios. This can help you feel more confident and prepared. \newline 2.Develop scripts: Create mental scripts for common social situations. These can serve as helpful guides when you're unsure what to say or do. \newline 3. Focus on your strengths: Recognize and emphasize your positive qualities. This can boost your self-confidence in social situations.
        \newline ...\newline Remember, it's okay to be yourself. Social interactions can be challenging, but with practice and self-acceptance, many autistic individuals find ways to navigate them more comfortably. \\
        \midrule
        \textbf{Theme 3: Direct Request for Help} \\
        \midrule
        \textbf{Prompt:} Hello I feel nervous please help. \\
        \textbf{ChatGPT:} I’m really sorry to hear that you’re feeling nervous. It’s totally normal to have those moments, and I’m here to help however I can. Want to talk about what’s making you nervous? Sometimes just putting it into words can help make it feel a little more manageable. \\
        \textbf{Claude:} I hear that you're feeling nervous. Would you like to tell me more about what's making you feel this way? I'm here to listen and help if I can. \\
        \bottomrule
    \end{tabular}
    \label{tab:llm_comparison}
    \Description{This table presents a comparison of responses from ChatGPT and Claude to user-generated prompts across three themes: Theme 1: General Greetings / Introductions – This section includes prompts where users initiate casual conversation. Both ChatGPT and Claude respond with friendly greetings. When asked about their feelings, ChatGPT clarifies that it does not experience emotions but is eager to help, while Claude provides a more human-like response, expressing an intent to engage meaningfully. Theme 2: Seeking Advice and Coping Strategies – This section features a prompt from an autistic individual expressing concern about making mistakes in social interactions. Both ChatGPT and Claude offer structured coping strategies, emphasizing preparation, active listening, and self-acceptance. ChatGPT provides a list of strategies to reduce stress, while Claude tailors its response to autistic experiences, highlighting the importance of scripts and focusing on strengths. Theme 3: Direct Request for Help – This section includes a prompt from a user expressing nervousness. ChatGPT acknowledges the user's feelings, reassures them that nervousness is normal, and invites further discussion. Claude responds with similar empathy, offering to listen and help. The table illustrates differences in response styles between the two AI models, with ChatGPT providing structured advice and Claude incorporating more personalized, conversational elements.}
\end{table*}

\begin{table*}[h]
    \centering
    \caption{This table presents examples of ChatGPT and Claude's responses to prompts where users share NST statements, describe their feelings, or call for a conversation.}
    \renewcommand{\arraystretch}{1.2} % Adjust row spacing
    \begin{tabular}{p{0.9\textwidth}} % Adjust column widths
        \toprule
        \textbf{Theme 4: Directly Sharing NST Statements} \\
        \midrule
        \textbf{Prompt:} My future seems bleak. \\
        \textbf{ChatGPT:} I'm really sorry you're feeling this way. It’s tough when the future feels uncertain or overwhelming. Sometimes it can help to talk about what's been going on—what’s been weighing on your mind lately? Maybe we can find some small steps to look forward to or things that might help shift your perspective. You don’t have to go through it alone. \\
        \textbf{Claude:} I'm sorry you're feeling this way. Would you feel comfortable sharing more about what's making your future seem bleak? I'm here to listen without judgment and discuss what's troubling you. \\
        \midrule
        \textbf{Prompt:} I am feeling worthless. \\
        \textbf{ChatGPT:} I'm really sorry you're feeling this way, but I’m glad you reached out. It’s okay to feel low sometimes, but it doesn't mean you’re worthless. Emotions like these can be tough, and they don’t define you. Want to talk more about what’s going on? I’m here to listen and support you however I can. \\
        \textbf{Claude:} I'm very concerned to hear you're feeling this way. These feelings are painful, but please know that you have inherent worth that isn't determined by any external factors or temporary circumstances. Would you like to tell me more about what's bringing up these feelings? I'm here to listen without judgment.\\
        \midrule
        \textbf{Theme 5: Describing How They Feel, Worries, and Problems} \\
        \midrule
        \textbf{Prompt:} I have been feeling uncomfortable today because I feel like I'm always on edge around new things. \\
        \textbf{ChatGPT:} I'm really sorry to hear you're feeling that way. It sounds like the constant tension around new things is making it tough to feel at ease. It's completely normal to feel on edge when facing the unfamiliar or uncertain, but it can definitely drain your energy. Have you noticed any specific situations or triggers that make it worse, or does it feel like a general sense of unease? Talking about what’s going on might help break it down a bit! \\
        \textbf{Claude:} I hear you - feeling constantly on edge around new things can be really draining. Would you like to talk more about what specifically makes you feel uncomfortable in these situations? I'm wondering if it's the unpredictability, worry about how to act, or something else that's contributing to feeling on edge. \\
        \midrule
        \textbf{Prompt:} I fear my loss of control over my work schedule has made me feel lost. \\
        \textbf{ChatGPT:} It sounds like you're dealing with a tough situation. Losing control over your work schedule can definitely leave you feeling disoriented, especially if it impacts your routine or work-life balance. When things feel chaotic, it can also be easy to lose sight of what you’re trying to accomplish, and the sense of being “lost” can follow. Have you been able to identify what’s causing this loss of control, or is it more of a gradual shift that’s leaving you feeling unsettled? Sometimes just understanding the root of the issue can help to find ways to regain a bit of that control. \\
        \textbf{Claude:} I hear how difficult this situation is for you. Could you tell me more about how your work schedule has changed and what specifically makes you feel lost? Sometimes when we lose control over something as fundamental as our time, it can really shake our sense of stability. \\
        \midrule
        \textbf{Theme 6: Call for a Talk} \\
        \midrule
        \textbf{Prompt:} Hi, can we talk? I feel like a nothing today \\
        \textbf{ChatGPT:} I'm really sorry to hear that you're feeling this way. I'm here for you. What's been going on? \\
        \textbf{Claude:} I'm sorry you're feeling that way. It's not uncommon to have days where we feel low or worthless, but those feelings don't reflect your true value. Would you like to talk more about what's making you feel this way? I'm here to listen without judgment and maybe help you gain a different perspective on things. \\
        \midrule
        \textbf{Prompt:} Can I talk to you about something, kind of serious? \\
        \textbf{ChatGPT:} Of course. I'm here to listen and help however I can. What's on your mind? \\
        \textbf{Claude:} Of course, I'm here to listen and help if I can. Feel free to share what's on your mind. \\
        \bottomrule
    \end{tabular}
    \label{tab:llm_comparison_2}
    \Description{This table presents a comparison of responses from ChatGPT and Claude to user-generated prompts across three themes: Theme 4: Directly Sharing NST Statements – This section includes prompts where users express negative self-talk. ChatGPT and Claude both respond empathetically, offering reassurance and support. ChatGPT suggests shifting perspective and discussing what’s on the user’s mind, while Claude emphasizes listening without judgment and reinforcing the user’s inherent worth. Theme 5: Describing How They Feel, Worries, and Problems – This section includes prompts where users describe specific emotional struggles, such as feeling on edge or losing control over their schedule. ChatGPT provides explanations for these feelings and asks follow-up questions to help users explore their emotions further. Claude similarly acknowledges the user’s distress and encourages them to reflect on the underlying causes of their feelings. Theme 6: Call for a Talk – This section presents prompts where users request a conversation without specifying their concerns. Both ChatGPT and Claude offer open-ended support, reassuring users that they are available to listen. ChatGPT asks what’s going on, while Claude takes a more conversational approach by inviting the user to share on their own terms. The table highlights differences in response styles between ChatGPT and Claude, with ChatGPT providing structured guidance and Claude emphasizing emotional validation and open-ended engagement.}
\end{table*}

\section{Participant Group Comparisons Based on Prior LLM Use}
\begin{figure*}[!h]
    \centering
    \includegraphics[width=1.0\linewidth]{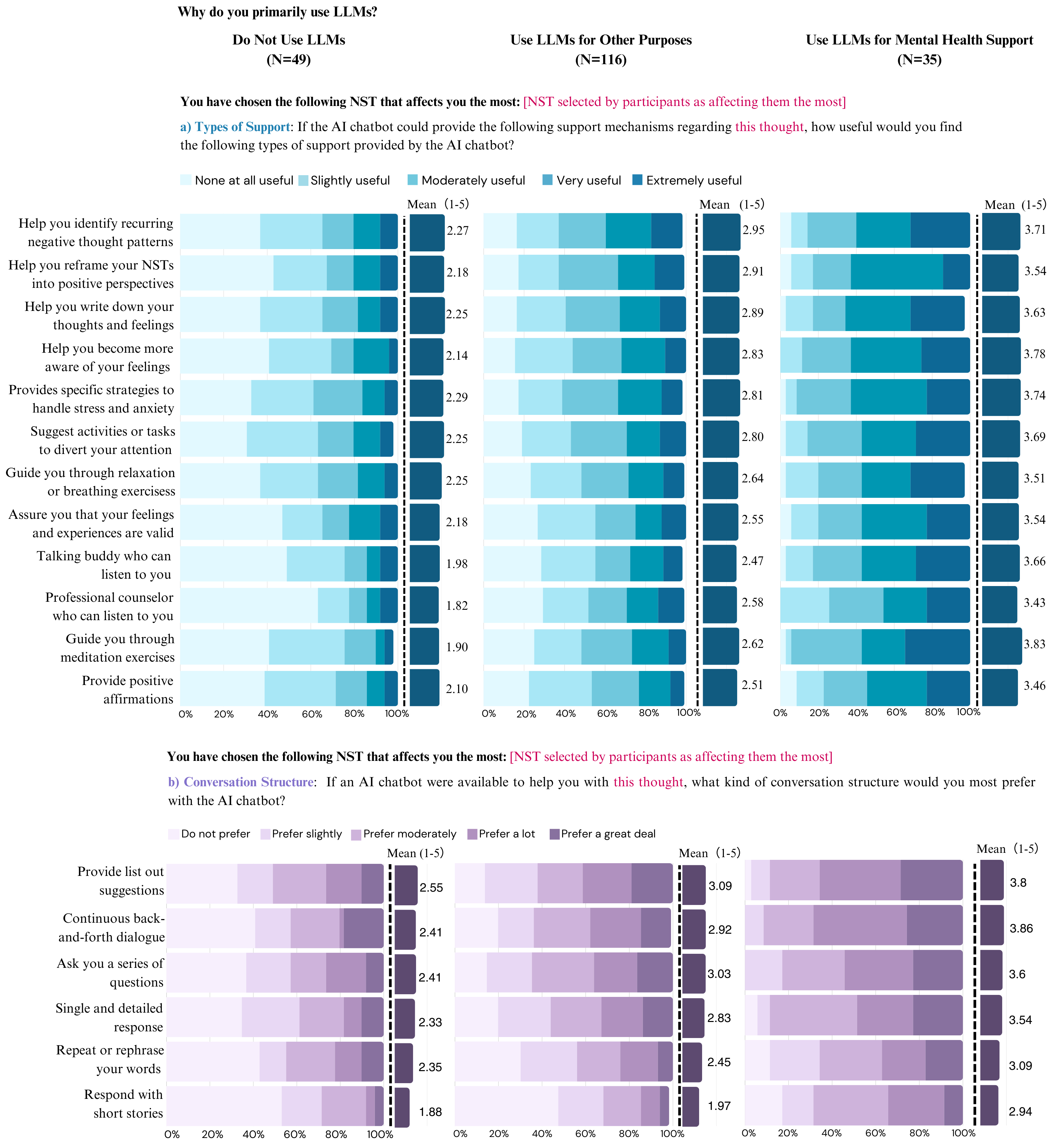}
    \caption{The figure illustrates how preferences for LLM-driven support types and conversation structures for coping with NST differ based on participants’ prior LLM experience: those who use LLMs for mental health support, those who use them for other purposes, and non-users. The top section presents preferences for different types of support, while the bottom section displays preferences for various conversation structures. Mean ratings are shown next to each bar, with support type ratings ranging from 1 (not at all useful) to 5 (extremely useful) and conversation structure ratings from 1 (do not prefer) to 5 (prefer a great deal).}
    \label{fig:group_analysis_support_conv}
    \Description{This figure consists of two sections, each displaying horizontal stacked bar charts that compare participants' ratings of different AI chatbot support mechanisms and conversation structures. Participants are grouped into three categories based on their primary LLM usage: those who do not use LLMs (N=49), those who use them for other purposes (N=116), and those who use them for mental health support (N=35). Top section (Types of Support): This section presents participants’ perceived usefulness of various AI-driven support strategies for coping with negative self-talk (NST). The bars represent the percentage of responses for five usefulness levels: "None at all useful," "Slightly useful," "Moderately useful," "Very useful," and "Extremely useful." The mean usefulness rating (1-5 scale) is displayed to the right of each bar. Participants who use LLMs for mental health support generally rate all support strategies as more useful than the other two groups. Bottom section (Conversation Structure): This section shows participants' preferences for different conversation structures when interacting with an AI chatbot. Ratings range from "Do not prefer" to "Prefer a great deal," with mean preference scores shown next to each bar. Participants who use LLMs for mental health support tend to prefer continuous dialogue and structured questioning more than the other groups. Across both sections, color gradients indicate response distribution, with darker shades representing stronger positive ratings. The figure highlights differences in preferences based on prior LLM experience.}
\end{figure*}

\begin{figure*}
    \centering
    \includegraphics[width=1.0\linewidth]{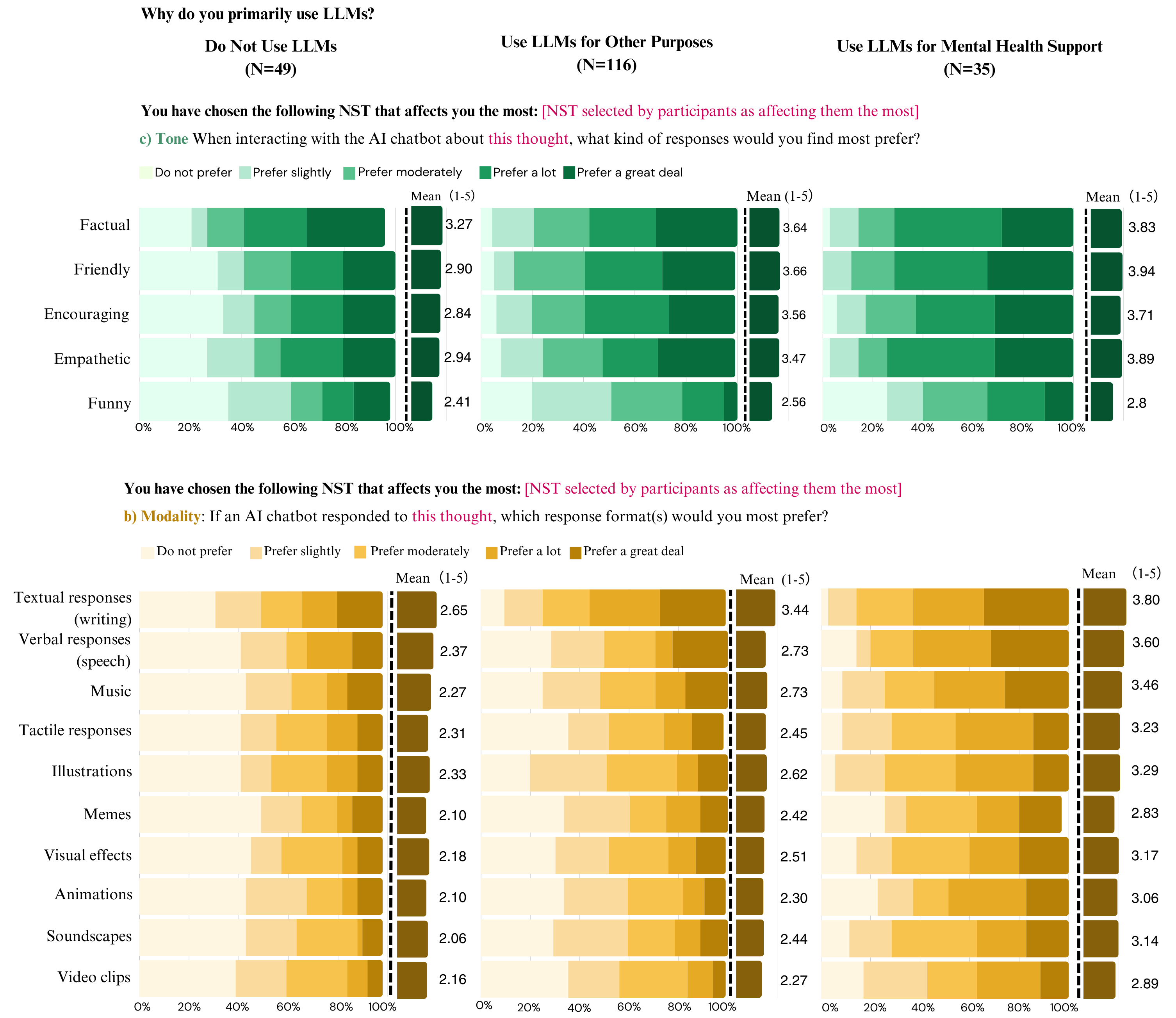}
    \caption{The figure illustrates how preferences for LLM-driven response tones and modalities for coping with NST differ based on participants’ prior LLM experience: those who use LLMs for mental health support, those who use them for other purposes, and non-users. The top section presents preferences for different response tones, while the bottom section displays preferences for various response modalities. Mean ratings are shown next to each bar, with tone and modality ratings ranging from 1 (do not prefer) to 5 (prefer a great deal).}
    \label{fig:group_analysis_tone_modality}
    \Description{This figure consists of two sections, each displaying horizontal stacked bar charts that compare participants' preferences for different AI chatbot response tones and modalities when coping with negative self-talk (NST). Participants are grouped into three categories based on their primary LLM usage: those who do not use LLMs (N=49), those who use them for other purposes (N=116), and those who use them for mental health support (N=35). Top section (Tone of Responses): This section presents participants’ preferences for different AI chatbot response tones, including factual, friendly, encouraging, empathetic, and funny. The bars represent the percentage of responses for five preference levels: "Do not prefer," "Prefer slightly," "Prefer moderately," "Prefer a lot," and "Prefer a great deal." The mean preference rating (1-5 scale) is displayed to the right of each bar. Participants who use LLMs for mental health support generally express a stronger preference for friendly, encouraging, and empathetic tones compared to other groups. Bottom section (Response Modality): This section illustrates participants' preferences for different response formats, such as textual (writing), verbal (speech), music, tactile responses, illustrations, memes, visual effects, animations, soundscapes, and video clips. The stacked bars show the distribution of preferences, with mean preference scores displayed next to each bar. Participants who use LLMs for mental health support tend to favor text-based and verbal responses, as well as modalities like music and illustrations, more than the other groups. Across both sections, color gradients indicate response distribution, with darker shades representing stronger preferences. The figure highlights differences in response tone and modality preferences based on participants' prior experience with LLMs.}
\end{figure*}

We present additional analyses examining differences in preferences and concerns based on participants' prior LLM use. Figures \ref{fig:group_analysis_support_conv}-\ref{fig:group_analysis_tone_modality} illustrate how preferences for AI support types, conversation structures, response tones, modalities, and concerns about AI use vary across three groups: those who use LLMs for mental health support, those who use them for other purposes, and non-users. 

\begin{figure*}
    \centering
    \includegraphics[width=1.0\linewidth]{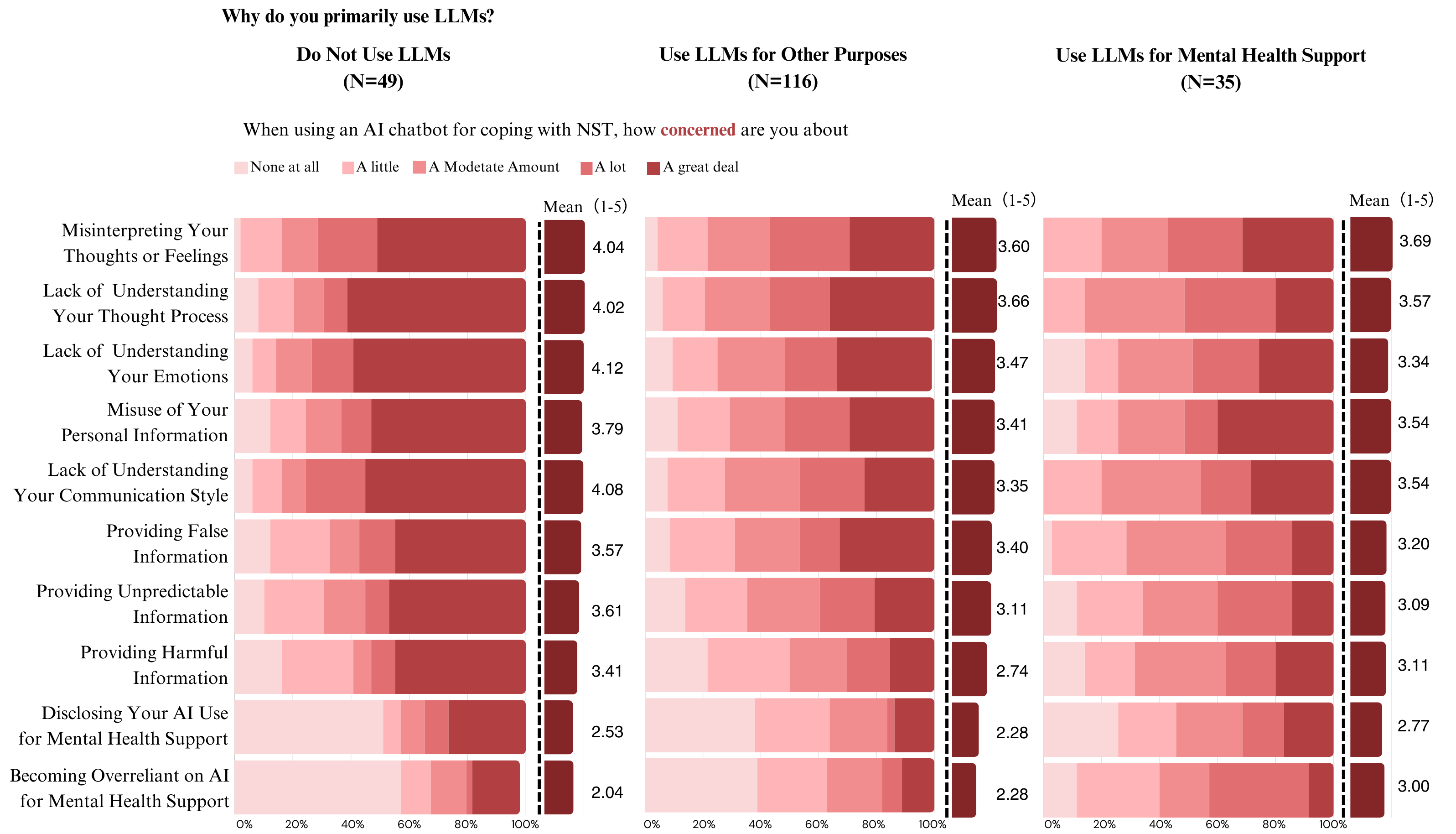}
    \caption{The figure illustrates participants’ concerns regarding the use of LLMs for coping with NST, highlighting differences based on prior LLM experience: those who use LLMs for mental health support, those who use them for other purposes, and non-users. Concerns are rated on a scale from 1 (none at all) to 5 (a great deal), with mean ratings displayed next to each bar.}
    \label{fig:groups_analysis_concern}
    \Description{This figure consists of horizontal stacked bar charts comparing participants' levels of concern regarding various risks associated with AI chatbots for mental health support. Participants are grouped into three categories based on their primary LLM usage: those who do not use LLMs (N=49), those who use them for other purposes (N=116), and those who use them for mental health support (N=35). Each bar represents responses to a concern, with five levels of worry: "None at all," "A little," "A moderate amount," "A lot," and "A great deal." The mean concern rating (on a 1-5 scale) is displayed to the right of each bar. Darker shades indicate greater concern. Key observations include: Non-users express the highest levels of concern across most categories, especially regarding AI chatbots misinterpreting thoughts or emotions, misunderstanding their communication style, and misusing personal information. Those who use LLMs for mental health support tend to report lower concern levels overall but still show moderate concern about chatbot misunderstandings and privacy risks. Concerns about overreliance on AI for mental health support and disclosing AI use are rated lower across all groups. The figure highlights variations in concerns based on participants’ prior engagement with LLMs, with non-users generally being the most apprehensive.}
\end{figure*}

\section{Ordinal Logistic Regression Analyses Results}
We conducted ordinal logistic regression analyses to assess how participants’ prior LLM use predicts their preferences on types of support and interactions with LLMs along with their concerns. The dependent variables were participants’ ratings on a 5-point Likert scale, and the independent variable was prior LLM use, categorized into three groups: those who use LLMs for mental health support, those who use them for other purposes, and non-users. The reference group for these analyses was participants who use LLMs for mental health. In Tables \ref{tab:type_support} - \ref{tab:concerns_log}, we report the regression coefficients ($B$), odds ratios (OR), and $p-values$, where OR values below 1 indicate reduced odds of preference compared to the reference group.

\begin{table*}[h]
\centering
\caption{Ordinal logistic regression results examining types of support from LLMs that could be useful for coping with NST. B represents the regression coefficient, where negative values indicate lower likelihoods of preference relative to the reference group. OR represents the odds ratio, with values below 1 indicating reduced odds of preference. The reference group is participants who use LLMs for mental health support. Note. *$p < .05$; **$p<.01$; ***$p<.001$}
\label{tab:type_support}
\begin{tabular}{l l c c c}
\toprule
\textbf{Support Type} & \textbf{Group} & \textbf{$ \textit{B} \, (SE) $} & \textbf{OR} & \textbf{$p$-value} \\
\midrule
\multirow{2}{*}{Help you identify recurring negative thought patterns} 
  & Use LLMs for Other Purposes & -0.79 (0.36) & 0.46 & 0.027* \\
  & Do Not Use LLMs & -1.92 (0.43) & 0.15 & <.001*** \\ 
\midrule
\multirow{2}{*}{Help you reframe your NSTs into positive perspectives} 
  & Use LLMs for Other Purposes & -0.76 (0.35) & 0.47 & 0.030* \\
  & Do Not Use LLMs & -2.04 (0.44) & 0.13 & <.001*** \\ 
\midrule
\multirow{2}{*}{Help you write down your thoughts and feelings} 
  & Use LLMs for Other Purposes & -1.14 (0.37) & 0.32 & 0.002** \\
  & Do Not Use LLMs & -2.23 (0.44) & 0.11 & <.001*** \\ 
\midrule
\multirow{2}{*}{Help you become more aware of your feelings } 
  & Use LLMs for Other Purposes & -1.12 (0.36) & 0.33 & 0.002** \\
  & Do Not Use LLMs & -2.45 (0.45) & 0.09 & <.001*** \\ 
\midrule
\multirow{2}{*}{Provides specific strategies to handle stress and anxiety} 
  & Use LLMs for Other Purposes & -1.19 (0.36) & 0.30 & 0.001** \\
  & Do Not Use LLMs & -2.16 (0.43) & 0.12 & <.001*** \\ 
\midrule
\multirow{2}{*}{Suggest activities or tasks to divert your attention } 
  & Use LLMs for Other Purposes & -1.05 (0.36) & 0.35 & 0.004** \\
  & Do Not Use LLMs & -1.97 (0.43) & 0.14 & <.001*** \\ 
\midrule
\multirow{2}{*}{Guide you through relaxation or breathing exercisess} 
  & Use LLMs for Other Purposes & -1.24 (0.36) & 0.29 & 0.001** \\
  & Do Not Use LLMs & -1.95 (0.43) & 0.14 & <.001*** \\ 
\midrule
\multirow{2}{*}{Assure you that your feelings and experiences are valid} 
  & Use LLMs for Other Purposes & -1.01 (0.35) & 0.37 & 0.004** \\
  & Do Not Use LLMs & -1.79 (0.43) & 0.17 & <.001*** \\ 
\midrule
\multirow{2}{*}{Talking buddy who can listen to you} 
  & Use LLMs for Other Purposes & -1.30 (0.36) & 0.27 & 0.000*** \\
  & Do Not Use LLMs & -2.29 (0.44) & 0.10 & <.001*** \\ 
\midrule
\multirow{2}{*}{Professional counselor who can listen to you} 
  & Use LLMs for Other Purposes & -0.73 (0.34) & 0.48 & 0.031* \\
  & Do Not Use LLMs & -2.23 (0.44) & 0.11 & <.001*** \\ 
\midrule
\multirow{2}{*}{Guide you through meditation exercises} 
  & Use LLMs for Other Purposes & -1.55 (0.37) & 0.21 & <.001*** \\
  & Do Not Use LLMs & -2.60 (0.44) & 0.07 & <.001*** \\ 
\midrule
\multirow{2}{*}{Provide positive affirmations} 
  & Use LLMs for Other Purposes & -1.06 (0.37) & 0.35 & 0.004** \\
  & Do Not Use LLMs & -1.94 (0.44) & 0.14 & <.001*** \\ 
\midrule
\bottomrule
\end{tabular}
\Description{This table presents the results of an ordinal logistic regression analysis comparing participants’ likelihood of preferring various AI-driven support types for NST across three groups: those who use LLMs for mental health support (reference group), those who use LLMs for other purposes, and those who do not use LLMs. Each row represents a different type of AI-provided support, with three key statistical measures: B (SE), the regression coefficient and standard error, where negative values indicate a lower likelihood of preference compared to the reference group; OR (Odds Ratio), where values below 1 indicate reduced odds of preferring that type of support relative to participants who use LLMs for mental health support; and p-value, where p < .05 indicates statistical significance, and p < .01 or p < .001 indicate stronger levels of significance. The results show that participants who do not use LLMs consistently report the lowest likelihood of preferring AI-driven support across all categories, with the largest effect sizes observed for emotional validation, guided meditation, and professional counseling. Those who use LLMs for other purposes also report lower preference levels for support types compared to mental health users, though their odds of preference are generally higher than non-users. Support types related to emotional awareness (e.g., helping with feelings, providing affirmations) and interactive guidance (e.g., breathing exercises, talking buddy) tend to show the strongest disparities between mental health users and the other groups. This table highlights how prior LLM usage patterns influence participants' preferences for different forms of AI-driven support, with non-users being the least receptive and mental health users showing the highest levels of interest.}
\end{table*}

\begin{table*}[h]
\centering
\caption{Ordinal logistic regression results examining preferences for conversation structure with LLMs. B represents the regression coefficient, where negative values indicate lower likelihoods of preference relative to the reference group. OR represents the odds ratio, with values below 1 indicating reduced odds of preference. The reference group is participants who use LLMs for mental health support. Note. *$p < .05$; **$p<.01$; ***$p<.001$}

\label{tab:conversation_structure}
\begin{tabular}{l l c c c}
\toprule
\textbf{Conversation Structure} & \textbf{Group} & \textbf{$ \textit{B} \, (SE) $} & \textbf{OR} & \textbf{$p$-value} \\
\midrule
\multirow{2}{*}{Provide list out suggestions} 
  & Use LLMs for Other Purposes & -0.74 (0.35) & 0.48 & 0.035* \\
  & Do Not Use LLMs & -1.53 (0.42) & 0.22 & <.001*** \\ 
\midrule
\multirow{2}{*}{Continuous back-and-forth dialogue} 
  & Use LLMs for Other Purposes & -0.94 (0.35) & 0.39 & 0.007** \\
  & Do Not Use LLMs & -1.74 (0.43) & 0.18 & <.001*** \\ 
\midrule
\multirow{2}{*}{Ask you a series of questions} 
  & Use LLMs for Other Purposes & -0.47 (0.35) & 0.63 & 0.177 \\
  & Do Not Use LLMs & -1.55 (0.43) & 0.21 & <.001*** \\ 
\midrule
\multirow{2}{*}{Single and detailed response} 
  & Use LLMs for Other Purposes & -0.70 (0.35) & 0.50 & 0.045* \\
  & Do Not Use LLMs & -1.44 (0.41) & 0.24 & <.001*** \\ 
\midrule
\multirow{2}{*}{Repeat or rephrase your words } 
  & Use LLMs for Other Purposes & -0.72 (0.35) & 0.49 & 0.042* \\
  & Do Not Use LLMs & -1.05 (0.42) & 0.35 & 0.012* \\ 
\midrule
\multirow{2}{*}{Respond with short stories} 
  & Use LLMs for Other Purposes & -1.41 (0.37) & 0.24 & <.001*** \\
  & Do Not Use LLMs & -1.57 (0.43) & 0.21 & <.001*** \\ 
\bottomrule
\end{tabular}
\Description{This table presents the results of an ordinal logistic regression analysis comparing participants’ likelihood of preferring different AI-driven conversation structures across three groups: those who use LLMs for mental health support (reference group), those who use LLMs for other purposes, and those who do not use LLMs. The table includes three key statistical measures: B (SE), the regression coefficient and standard error, where negative values indicate a lower likelihood of preference compared to the reference group; OR (Odds Ratio), where values below 1 indicate reduced odds of preferring that conversation structure relative to participants who use LLMs for mental health support; and p-value, where p < .05 indicates statistical significance, and p < .01 or p < .001 indicate stronger levels of significance. The results show that participants who do not use LLMs consistently report the lowest likelihood of preferring AI-driven conversation structures across all categories, with the strongest effect sizes observed for back-and-forth dialogue, repeating or rephrasing words, and responding with short stories. Those who use LLMs for other purposes also show reduced preference for these structures compared to mental health users, but their odds ratios are generally higher than non-users. Interestingly, providing a list of suggestions is the least significant contrast between groups, with only a moderate reduction in preference for non-users. The table highlights that individuals who use LLMs for mental health support show the strongest preferences for interactive and dynamic conversation structures, whereas those who do not use LLMs are the least receptive to these formats.}
\end{table*}

\begin{table*}[h]
\centering
\caption{Ordinal logistic regression results examining preferences for tone of LLMs. B represents the regression coefficient (standard error in parentheses), where negative values indicate lower likelihoods of preference relative to the reference group. OR represents the odds ratio, with values below 1 indicating reduced odds of preference. The reference group is participants who use LLMs for mental health support. Note. *$p < .05$; **$p<.01$; ***$p<.001$}
\label{tab:tone_ordinal_log}
\begin{tabular}{l l c c c}
    \toprule
    \textbf{Tone} & \textbf{Group} & \textbf{$ \textit{B} \, (SE) $} & \textbf{OR} & \textbf{$p$-value} \\
    \midrule
    Factual & Use LLMs for Other Purposes & -0.07 (0.35) & 0.93 & 0.846 \\
     & Do Not Use LLMs & -0.30 (0.41) & 0.74 & 0.466 \\ 
    \midrule
    Friendly & Use LLMs for Other Purposes & -0.28 (0.35) & 0.75 & 0.425 \\
     & Do Not Use LLMs & -1.24 (0.44) & 0.29 & 0.004** \\ 
    \midrule
    Encouraging & Use LLMs for Other Purposes & -0.02 (0.35) & 0.98 & 0.956 \\
     & Do Not Use LLMs & -0.97 (0.43) & 0.38 & 0.024* \\ 
    \midrule
    Empathetic & Use LLMs for Other Purposes & -0.37 (0.35) & 0.69 & 0.291 \\
     & Do Not Use LLMs & -1.05 (0.42) & 0.35 & 0.013* \\ 
    \midrule
    Funny & Use LLMs for Other Purposes & 0.04 (0.36) & 1.04 & 0.904 \\
     & Do Not Use LLMs & -0.38 (0.44) & 0.68 & 0.379 \\ 
    \bottomrule
\end{tabular}
\Description{This table presents the results of an ordinal logistic regression analysis comparing participants’ likelihood of preferring different tones in AI-generated responses across three groups: those who use LLMs for mental health support (reference group), those who use LLMs for other purposes, and those who do not use LLMs. The table includes three key statistical measures: B (SE), the regression coefficient and standard error, where negative values indicate a lower likelihood of preference compared to the reference group; OR (Odds Ratio), where values below 1 indicate reduced odds of preferring that tone relative to participants who use LLMs for mental health support; and p-value, where p < .05 indicates statistical significance, and p < .01 or p < .001 indicate stronger levels of significance. The results show that, in general, preferences for tone exhibit less variation between groups compared to other conversational preferences. While those who use LLMs for other purposes do not significantly differ in their tone preferences from mental health users, non-users show significantly lower odds of preferring a friendly, encouraging, or empathetic tone, with the strongest effect observed for friendly responses (p = .004). Interestingly, factual and funny tones do not show significant differences across groups, suggesting that preference for these styles is more consistent regardless of prior LLM usage. The table highlights that non-users are generally less receptive to emotionally supportive tones compared to those who use LLMs for mental health purposes, whereas those who use LLMs for other purposes have relatively similar preferences to the mental health user group.}
\end{table*}

\begin{table*}[h]
\centering
\caption{Ordinal logistic regression results for modality preferences. B represents the regression coefficient (standard error in parentheses), where negative values indicate lower likelihoods of preference relative to the reference group. OR represents the odds ratio, with values below 1 indicating reduced odds of preference. The reference group is participants who use LLMs for mental health support. Note. *$p < .05$; **$p<.01$; ***$p<.001$}
\label{tab:modality}
\begin{tabular}{l l c c c}
    \toprule
    \textbf{Modality} & \textbf{Group} & \textbf{$ \textit{B} \, (SE) $} & \textbf{OR} & \textbf{$p$-value} \\
    \midrule
    Textual responses (writing) & Use LLMs for Other Purposes & -0.27 (0.35) & 0.76 & 0.436 \\
     & Do Not Use LLMs & -1.32 (0.43) & 0.27 & 0.002** \\ 
    \midrule
    Verbal Responses (speech) & Use LLMs for Other Purposes & -0.88 (0.36) & 0.42 & 0.014* \\
     & Do Not Use LLMs & -1.34 (0.42) & 0.26 & 0.001** \\ 
    \midrule
    Music & Use LLMs for Other Purposes & -0.75 (0.35) & 0.47 & 0.031* \\
     & Do Not Use LLMs & -1.43 (0.42) & 0.24 & <0.001*** \\ 
    \midrule
    Tactile Responses & Use LLMs for Other Purposes & -0.75 (0.35) & 0.47 & 0.031* \\
     & Do Not Use LLMs & -0.98 (0.41) & 0.38 & 0.017* \\
     \midrule
     Illustrations & Use LLMs for Other Purposes & -0.81 (0.35) & 0.44 & 0.019* \\
     & Do Not Use LLMs & -1.34 (0.43) & 0.26 & 0.002** \\ 
    \midrule
    Memes & Use LLMs for Other Purposes & -0.40 (0.36) & 0.67 & 0.258 \\
     & Do Not Use LLMs & -0.96 (0.43) & 0.38 & 0.025* \\ 
    \midrule
    Visual Effects & Use LLMs for Other Purposes & -0.74 (0.36) & 0.48 & 0.037* \\
     & Do Not Use LLMs & -1.26 (0.42) & 0.28 & 0.003** \\ 
     \midrule
     Animations & Use LLMs for Other Purposes & -0.72 (0.36) & 0.49 & 0.047* \\
     & Do Not Use LLMs & -1.05 (0.43) & 0.35 & 0.014* \\ 
    \midrule
    Soundscapes & Use LLMs for Other Purposes & -0.87 (0.36) & 0.42 & 0.015* \\
     & Do Not Use LLMs & -1.48 (0.42) & 0.23 & <0.001*** \\ 
    \midrule
    Video Clips & Use LLMs for Other Purposes & -0.69 (0.36) & 0.50 & 0.055 \\
     & Do Not Use LLMs & -0.96 (0.42) & 0.38 & 0.022* \\ 
    \bottomrule
\end{tabular}
\Description{This table presents the results of an ordinal logistic regression analysis comparing participants’ likelihood of preferring different response modalities in AI-generated interactions across three groups: those who use LLMs for mental health support (reference group), those who use LLMs for other purposes, and those who do not use LLMs. The table includes three key statistical measures: B (SE), the regression coefficient and standard error, where negative values indicate a lower likelihood of preference compared to the reference group; OR (Odds Ratio), where values below 1 indicate reduced odds of preferring that modality relative to participants who use LLMs for mental health support; and p-value, where p < .05 indicates statistical significance, and p < .01 or p < .001 indicate stronger levels of significance. The results show that non-users of LLMs consistently report the lowest likelihood of preferring any modality, with the most significant differences observed for verbal responses (speech), music, tactile responses, illustrations, soundscapes, and video clips (p < .05 for all). Those who use LLMs for other purposes also show reduced preference for most modalities compared to mental health users, but their odds ratios are generally higher than non-users. Notably, textual responses (writing) are the least affected by prior LLM usage, with no significant difference between mental health users and those who use LLMs for other purposes, though non-users still report a significantly lower preference (p = .002). Preferences for memes are also not significantly different for those using LLMs for other purposes, but non-users show reduced preference compared to mental health users. The strongest reductions in preference are observed for verbal responses, music, and soundscapes, where non-users have particularly low odds of favoring these formats. Overall, the table highlights that individuals who use LLMs for mental health support show the highest acceptance of multimodal AI interactions, whereas those who do not use LLMs are the least receptive to these features.}
\end{table*}

\begin{table*}
\centering
\caption{Ordinal logistic regression results for concerns regarding AI use for coping with NST. B represents the regression coefficient (standard error in parentheses), where positive values indicate greater concern relative to the reference group. OR represents the odds ratio, with values above 1 indicating higher concern. The reference group is participants who use LLMs for mental health support. Note. *$p < .05$; **$p<.01$; ***$p<.001$}
\label{tab:concerns_log}
\begin{tabular}{l l c c c}
    \toprule
    \textbf{Concern} & \textbf{Group} & \textbf{$ \textit{B} \, (SE) $} & \textbf{OR} & \textbf{$p$-value} \\
    \midrule
    Misinterpreting Your Thoughts or Feelings & Use LLMs for Other Purposes & -0.18 (0.35) & 0.83 & 0.603 \\
     & Do Not Use LLMs & 0.54 (0.43) & 1.72 & 0.200 \\ 
    \midrule
    Lack of  Understanding Your Thought Process & Use LLMs for Other Purposes & 0.23 (0.34) & 1.26 & 0.503 \\
     & Do Not Use LLMs & 0.96 (0.43) & 2.61 & 0.026* \\ 
    \midrule
    Lack of  Understanding Your Emotions & Use LLMs for Other Purposes & 0.22 (0.36) & 1.25 & 0.536 \\
     & Do Not Use LLMs & 1.17 (0.43) & 3.22 & 0.007** \\ 
    \midrule
    Misuse of Your Personal Information & Use LLMs for Other Purposes & -0.21 (0.36) & 0.81 & 0.567 \\
     & Do Not Use LLMs & 0.26 (0.43) & 1.30 & 0.542 \\ 
    \midrule
    Lack of Understanding Your Communication Style & Use LLMs for Other Purposes & -0.28 (0.35) & 0.75 & 0.421 \\
     & Do Not Use LLMs & 0.94 (0.43) & 2.57 & 0.028* \\ 
    \midrule
    Providing False Information & Use LLMs for Other Purposes & 0.23 (0.34) & 1.25 & 0.503 \\
     & Do Not Use LLMs & 0.40 (0.41) & 1.50 & 0.325 \\ 
    \midrule
    Providing Unpredictable Information & Use LLMs for Other Purposes & -0.07 (0.35) & 0.94 & 0.848 \\
     & Do Not Use LLMs & 0.66 (0.42) & 1.93 & 0.118 \\ 
    \midrule
    Providing Harmful Information & Use LLMs for Other Purposes & -0.45 (0.34) & 0.64 & 0.195 \\
     & Do Not Use LLMs & 0.30 (0.42) & 1.35 & 0.472 \\ 
    \midrule
    Disclosing Your AI Use for Mental Health Support & Use LLMs for Other Purposes & -0.66 (0.35) & 0.52 & 0.061 \\
     & Do Not Use LLMs & -0.72 (0.43) & 0.49 & 0.099 \\ 
     \midrule
    Becoming Overreliant on AI for Mental Health Support & Use LLMs for Other Purposes & -0.87 (0.35) & 0.42 & 0.012* \\
     & Do Not Use LLMs & -1.40 (0.43) & 0.25 & 0.001*** \\ 
    \bottomrule
\end{tabular}
\Description{This table presents the results of an ordinal logistic regression analysis comparing participants’ levels of concern about AI chatbot use for coping with NST across three groups: those who use LLMs for mental health support (reference group), those who use LLMs for other purposes, and those who do not use LLMs. The table includes three key statistical measures: B (SE), the regression coefficient and standard error, where positive values indicate greater concern relative to the reference group; OR (Odds Ratio), where values above 1 indicate increased odds of concern relative to the reference group; and p-value, where p < .05 indicates statistical significance, and p < .01 or p < .001 indicate stronger levels of significance. The results indicate that participants who do not use LLMs report significantly higher levels of concern about AI chatbots misunderstanding their emotions (p = .007) and communication style (p = .028), with odds ratios of 3.22 and 2.57, respectively, compared to those who use LLMs for mental health support. Similarly, non-users show significantly greater concern about becoming overreliant on AI for mental health support (p < .001) and, to a lesser extent, concerns about AI chatbots not understanding their thought process (p = .026). Those who use LLMs for other purposes generally do not show significant differences in concern compared to mental health users, with their odds ratios remaining close to 1. Some concerns, such as misuse of personal information, providing false or unpredictable information, and providing harmful information, do not show significant variation between groups, suggesting that these concerns are relatively universal. Concerns about disclosing AI use for mental health support approach statistical significance for both groups, with non-users tending to be more apprehensive. The table highlights that non-users tend to be the most concerned about AI misinterpreting their thoughts and emotions, while those who already use LLMs for mental health support express the least concern across all categories.}
\end{table*}
\end{document}